\documentclass{ieeeaccess}
\usepackage{cite}
\usepackage{amsmath,amssymb,amsfonts}
\usepackage{algorithmic}
\usepackage{graphicx}
\usepackage{textcomp}
\def\BibTeX{{\rm B\kern-.05em{\sc i\kern-.025em            b}\kern-.08em
		T\kern-.1667em\lower.7ex\hbox{E}\kern-.125emX}}

\usepackage{amssymb}
\usepackage{amsthm, amsmath}

\usepackage{booktabs}
\usepackage{xcolor}
\usepackage{tabularx}
\usepackage{adjustbox}
\usepackage{textcomp}

\usepackage{subcaption}

\usepackage{dcolumn}
\usepackage{bm}
\usepackage{braket}

\PassOptionsToPackage{hyphens}{url}\usepackage[colorlinks=true,urlcolor=blue,citecolor=blue,linkcolor=blue]{hyperref}

\usepackage{soul}
\usepackage{amsmath}
\usepackage{algorithm}
\usepackage{algorithmic}
\usepackage{physics}
\newtheorem{assumption}{Assumption}

\DeclareMathOperator*{\argmin}{arg\,min}

\newcommand{\threeadmm}{$3$-ADMM-H}
\newcommand{\transp}{\intercal}
\newcommand{\mathds}{}
\newcommand{\nz}{\mathrm{nz}}
\newcommand{\ru}[1]{#1}

\newcommand{\rev}[1]{{#1}}

\begin{document}
	\history{
		Received August 10, 2020; revised September 28, 2020; accepted October 7, 2020; date of publication October 13, 2020;
date of current version November 5, 2020.
	}
	\doi{10.1109/TQE.2020.3030314}

\title{Quantum Computing for Finance: State of the Art and Future Prospects}
\author{\uppercase{Daniel J. Egger \authorrefmark{1}, Claudio Gambella \authorrefmark{1}, Jakub Marecek \authorrefmark{1}, Scott McFaddin \authorrefmark{1}, Martin Mevissen \authorrefmark{1}, Rudy Raymond \authorrefmark{1}}, 
	\uppercase{Andrea Simonetto \authorrefmark{1}},
		\uppercase{Stefan Woerner \authorrefmark{1}, Elena Yndurain \authorrefmark{1}}
}
\address[1]{IBM Quantum}
\tfootnote{
Jakub Marecek, who has performed the present work while at IBM, is currently with the Department of Computer Science, Faculty of
Electrical Engineering, Czech Technical University, 121 35, Prague, Czech Republic.
}

\markboth
{D.~J. Egger \headeretal: Quantum Computing for Finance}
{D.~J. Egger \headeretal: Quantum Computing for Finance}

\corresp{Corresponding author: Andrea Simonetto (email: andrea.simonetto@ibm.com).}

\begin{abstract}
This paper outlines our point of view regarding the applicability, state of the art, and potential of quantum computing for problems in finance. We provide an introduction to quantum computing as well as a survey on problem classes in finance that are computationally challenging classically and for which quantum computing algorithms are promising. In the main part, we describe in detail quantum algorithms for specific applications arising in financial services, such as those involving simulation, optimization, and machine learning problems. In addition, we  include demonstrations of quantum algorithms on IBM Quantum back-ends and discuss the potential benefits of quantum algorithms for problems in financial services. We conclude with a summary of technical challenges and future prospects. 
\end{abstract}


\titlepgskip=-15pt

\maketitle

\section{Introduction}
\label{intro}

In the financial services industry there are many computationally challenging problems arising in applications across asset management, investment banking and retail \& corporate banking. Quantum computing holds the promise of revolutionizing how we solve such computationally challenging problems. With the first, noisy quantum devices - leveraging the principles of quantum mechanics - available publicly today \rev{(see e.g.,~\cite{LaRose2019,Wiki2020})}, the applicability of quantum computing for problems in finance and demonstrating Quantum Advantage in first applications are active topics of current research \rev{\cite{Orus2019, BGK2018, Egger2019, Rebentrost_2018, Stamatopoulos_2020, Suzuki2019, Woerner2019, braine2019quantum, gambella2020multi, nannicini2019performance}}. 

In this paper, we provide an introduction to quantum computing and the necessary foundational concepts to understand this new technology and its implications to the financial services industry \cite{Nielsen2000}. We extend previous summaries \cite{Orus2019, ibmibv1, ibmibv2} in multiple directions, as follows.
First, we review the main algorithms, the benefits they bring as well as the technical challenges they pose, and how to approach problems from a quantum perspective. We then also highlight the economic benefits that applying quantum computing may bring to financial institutions in improving operations, revenues, and quality. Algorithms are categorized based on the type of problems they solve and mapped to the financial solutions they can be applied to. We showcase real-life examples of using quantum computing algorithms, explaining how the problems are solved and the solutions obtained. Overall, we offer a holistic, practical guide to quantum computing and its applicability to financial problems for financial institutions in Banking, Financial Markets and Insurance.

All computing systems rely on a fundamental ability to store and manipulate information. Today's classical computers manipulate individual bits, which store information as binary 0 and 1 states. Millions of bits work together to process and display information with a \textit{speed} that everyone is familiar with on smart phones, laptops, and the servers in the cloud. Quantum computers use the physical phenomena of nature to manipulate information via quantum mechanics. At this fundamental level we have quantum bits, or qubits.
Unlike a bit that has to be a 0 or a 1 (or, their probabilistic combination), a qubit can be in a complex-value-weighted combination of states called a \textit{superposition}.
Multiple qubits may be purposefully \textit{entangled} into linear combinations of (complex-valued-weighted) states across the qubits, which "correlates" them so their quantum state cannot be described independently, i.e., entangled states. \rev{And quantum \textit{interference} allows us to bias the measurement of a qubit toward a desired state, thus, controlling the probability a system of qubits collapses into particular measurement states.}
Quantum computers are not a replacement for classical ones. They complement the traditional systems by possibly being able to solve some forms of intractable problems that \textit{blow up}, or become extremely large or time-consuming during computation. Remarkable progress in the control and construction of quantum computing hardware in recent years has led to the development of systems with 10's of physical qubits, that in the absence of quantum error correction are dubbed noisy quantum processors\rev{~\cite{Preskill2018}}.  
This has sparked strong interest in the pursuit of Quantum Advantage -- the exploration of computational tasks \rev{which a quantum computer solves faster than a classical one. An example of a task for which Quantum Advantage has been proven was introduced in \cite{BGK2018}} 
This is also a step on the way to the development of Fault-Tolerant Universal Quantum Computers (FTQC), which require error correction \rev{\cite{gottesman1997stabilizer, aharonov2008fault, campbell2017roads}} and which will allow for arbitrary quantum algorithms backed by theory \rev{that proves} quantum speedup compared to classical algorithms.

The quantum computing hardware pursued by IBM involves superconducting quantum circuits\rev{~\cite{ibmsystem2020}}. The fundamental building blocks of the hardware are Josephson junction-based qubits called transmons. When cooled to ca.~10mK these transmons behave as artificial atoms, where the two lowest energy levels may be employed as the computational 0 and 1 states. Over the last two decades, progress with superconducting qubit technology has been driven by tremendous improvements in coherence, control and fabrication capabilities. A metric that has been proposed and employed by IBM to measure progress in development in quantum computing is Quantum Volume ~\cite{Cross2019}. Quantum Volume indicates the relative complexity of a problem that can be solved on a quantum computer, and it depends on a number of factors such as number of qubits, coherence time, measurement errors, device cross talk, circuit compiler efficiency, and others. IBM currently has \rev{more than 20} superconducting processors available via the cloud with up to \rev{65} qubits, and \rev{put forward a roadmap for scaling quantum technology in the years ahead \cite{Gambetta2020_roadmap}.}

In the near-term future, quantum computers will continue to rely on noisy qubits with relatively high error rates and limited coherence times. In this era of noisy quantum devices\rev{~\cite{Preskill2018}}, \rev{we will be confidently in the realm of} Quantum Advantage once we are able to solve a \rev{good number of significant} real-world problems more efficiently with the help of quantum devices than compared with solving them on a classical computers only. The problem classes where we expect to demonstrate advantage first include (i) modelling physical processes of nature, \rev{(ii) obtaining better solutions to optimization problems, and (iii) finding better patterns within machine learning processes, since first promising proposals for near-term compatible quantum heuristics exist \cite{peruzzo2014variational, farhi2014quantum, Havl_ek_2019}}.

At a basic level, any solution built on a quantum computer will comprise three fundamental steps:

\paragraph{The Loading Step} The solution must load its data from a classical computer into the quantum computer in a small number of steps.
This results in a superposition which combines the fundamental states described above in a way that reflects the data set being loaded. While loading data into a computer is often ignored as a triviality in classical computing, quantum data loading can be a substantial aspect of solution design.
While there are loading techniques which are  linear in the size of the data set, this can often eclipse the coherence times of the physical superposition,
and techniques are often employed to reduce the input data before loading or encode the data into the computational steps.

\paragraph{The Compute Step} Once data is loaded, the solution must rapidly perform a computation on the loaded data within the quantum computer.
This involves manipulation of the qubits in a manner that changes the fundamental states in a way that reflects the outcome of a desired computation.
There is an increasing body of research into quantum algorithms that solve problems that are considered computationally difficult or intractable classically.
Quantum computations often compute their results as an approximation to an optimal value within a high dimensional search space,
and often exploit the nature of quantum superpositions to simultaneously consider vast numbers of possibilities.
The computed "output" superposition reflects a probabilistic distribution of possible outcomes, with the preferred outcomes associated with higher probabilities in the distribution.

\paragraph{The Measurement Step} The measurement step makes an observation of the computed "output" superposition and reports it back to a classical computer.
However, the act of observation ``snaps'' the superposition back to a basis state, so the computation step must be designed in such as way that the solution is often encoded in a narrow decision space.
Quantum algorithms are often repeated multiple times, with each repetition called a \textit{shot}.
Each shot reports an output among the distribution of possible outputs from the compute step.
With multiple repetitions, a probabilistic picture emerges of which output has received the highest probability within the superposition,
and an ``answer'' may be read.
Typically hundreds or thousands of shots are used.
It should be noted that the higher the number of qubits, the higher requirements on precision, and hence the higher the requirements on the number of measurements.

\section{Problems in Financial Services}

Financial services are a forward-looking industry that has always been in the lookout to leverage new technologies to increase profits. Broadly, 
this industry covers three vertical sectors (c.f. also
Table \ref{tab0} for segments): 
\begin{itemize}
    \item 
    \textbf{Banking:} Banking products are mainly bank accounts, investments, loans for commercial and retail customers. Their main challenges are to balance cash with interest rates, while controlling threats related to liquidity, fraud, money laundry, or non-performing loans.    
    \item
    \textbf{Financial Markets:} Focused mainly on future gains and the marketplace to sell and buy assets by dealers, exchanges, brokers, or clearing houses. Their main challenges are to manage geographic time-zones, immediacy needs, and counter-party risk. 
    \item 
    \textbf{Insurance:} Health insurance, automobile \& property, life insurance \& annuity and re-insurance. The main challenge here is to maximize premiums and manage threats related to unplanned risks such as catastrophes or market crashes.
\end{itemize}

\begin{table}
\caption{Segments of the financial services industry} 
\sf
\small
\begin{tabularx}{\columnwidth}{lll}
\toprule
Banking & Financial Markets & Insurance \\
\toprule
Monetary Authorities & Commodities & Health Insurance \\ 
Retail Banking & Stock Exchanges & Property \& \\
&& Casualty Insurance \\ Commercial & Bond Markets & Life Insurance \\
Banking && \& Annuity \\
Investment Banking & Money Markets & Reinsurance \\
Non-Banking FIs & Derivatives\\
\bottomrule
\end{tabularx}
\label{tab0}
\end{table}

The digital financial services revolution is full blast disrupting the industry and opening the door to new players threatening the current status quo \rev{\cite{pwc_2020_technology}}: FinTechs and InsurTechs with new digital-style offerings,  RegTech with automatized regulation processes, and new competition from other industry corporations that can now offer digital financial services to their customers. In addition, clients are demanding customized offerings based on their behavioral data, for example, personalized insurance premiums based on their life-events or targeted loan offers for smaller customer segments. The regulatory environment poses an operational challenge by requiring risk mitigation and strict compliance. 
All these trends have created new financial services experiences: Smooth global supply-chains for trade finance asset management, AI augmentation with trusted decision making for investment banking, and banking platforms that become finance `as-a-service'. 

In this paper, we consider the regulatory framework given by the Basel III rules, which is being implemented by regulators world-wide. 
Under Basel III, a key performance indicator for both the regulator and regulated entities is capital adequacy ratio (CAR), which is a ratio of the combined tier-1 and tier-2 capital and risk-weighted holdings. In particular, the minimum is 8\%, while the required ratio is 10.5\%. 

Moreover, the Liquidity Coverage Ratio (LCR) reform \cite{basedIIIlcr}, with the aim of improving short-term resilience,  promotes the holding of unencumbered high-quality liquid assets (HQLA), whose amounts are tested in the so-called 30 calendar day liquidity stress scenario. By January 1st, 2019, the LCR, i.e., the proportion of the value of the stock of HQLA in stressed conditions to total net cash outflows in the same scenario, has risen to 100\%. 
Within HQLA, ``Level 1'' assets include cash and certain state-backed securities, as well as select other safe assets.
``Level 2'' include further state-backed securities and bonds of non-state and non-bank entities with long-term rating AA- or better.
``Level 2'' assets can only comprise up to 40\% of the stock of HQLA. 
Further, the same LCR reform introduced diversification requirements on the stock of HQLA.
These extensions complicate both risk assessment and portfolio management problems considerably.

\begin{figure}[t!]
	\centering
		\includegraphics[width=\columnwidth]{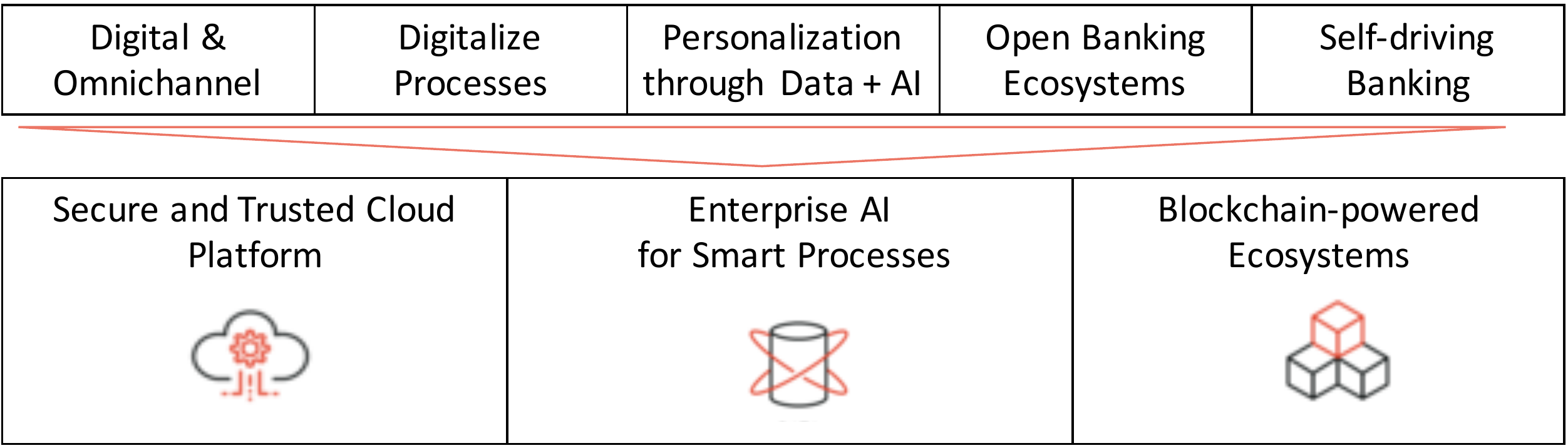}
		\caption{Transformative technologies in support of digital themes in financial services \cite{Bhaskaran2019}.}
	\label{fig:TransformativeTech}
\end{figure}

\begin{figure*}[t!]
	\centering
		\includegraphics[width=0.8\textwidth]{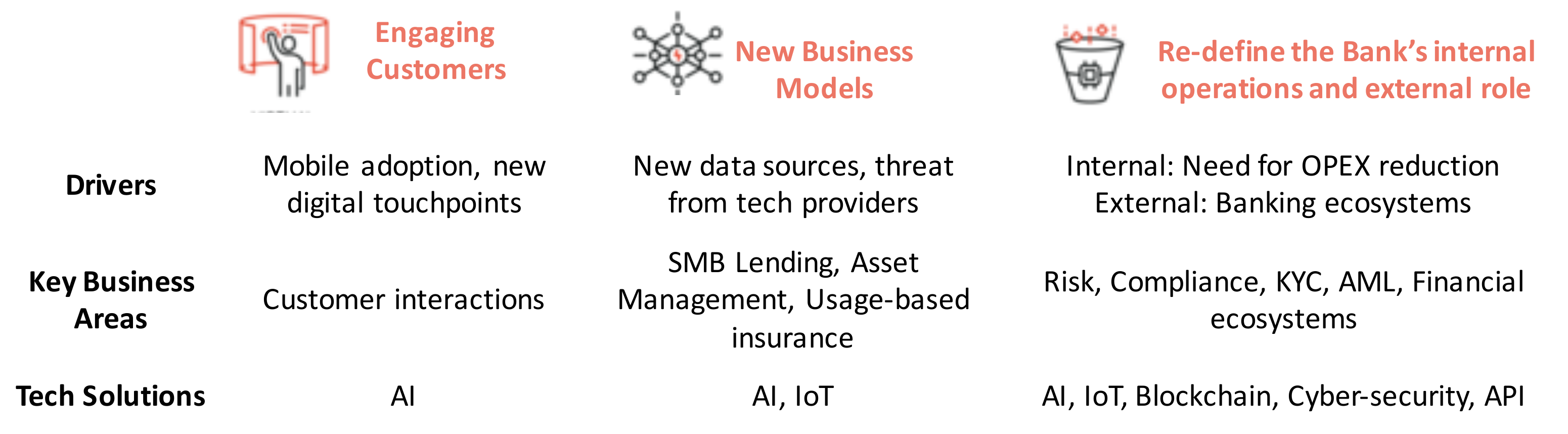}
		\caption{Finance industry transformation in waves driving to a digital future \cite{Bhaskaran2019}.}
	\label{fig:TransformationWaves}
\end{figure*}

Indeed, many problems, such as risk management and portfolio management, rely on various risk measures that we now introduce for future reference.
Volatility captures the risk in a portfolio of assets.
It is the standard deviation of the returns which can be calculated from the variance $\omega^{\intercal}\Sigma\omega$.
Here, $\omega$ is a (column) vector made of the weights of each asset in the portfolio and $\Sigma$ is the covariance matrix of the returns of the assets.

Risk management often uses Value at Risk (VaR) as a risk measure.
$\rm{VaR}_\alpha$ of a random variate $X$ is the $1-\alpha$ quantile of the loss distribution $Y=-X$. 
VaR is, therefore, the minimal $\gamma$ such that the probability that $X$ exceeds $\gamma$ is $\alpha$, i.e.,
\begin{align} 
	{\rm VaR}_\alpha(X) :=
	-{\inf}\{\gamma~{\rm such~that}~F_X(\gamma)>\alpha\}, \label{eqn:var}
\end{align}
where $F_X(x)$ is the cumulative distribution function of $X$.
Since VaR is a quantile it has the short-coming that it is not sensitive to extreme losses in the tail of $X$.
The conditional value at risk CVaR is, therefore, often used as an additional risk metric.
$\rm{CVaR}_\alpha$, sometimes also called expected shortfall, is the expectation value of all losses up to the $\rm{VaR}_\alpha$.
In the financial services challenges discussed in the remainder of this paper, we are assuming a market environment operating under the Basel III regulations.

{ Planned updates to the framework, upcoming under Basel IV, focus on the approach to calculate Risk weighted assets (RWA) regardless of risk type. Banks will need to change their projection models towards forward-looking statements which are statements that are not solely based on historical facts and therefore have more assumptions. This will require more scenario building to comply with the new requirements for Capital lower limits (72.5\% ``output floor''), credit risk with common approach rather than internal, market risk sensitivity-driven analysis as a standard, and operational risk measured by  ``unadjusted business indicator'' leveraging historical loss data. Overall Basel IV will reshape banks' trading activities and portfolio structures. 
}

In this paper, the focus is in three areas of financial services, where problems challenging for classical computers arise today:
\begin{enumerate}
    \item Asset Management;
    \item Investment Banking;
    \item Retail and Corporate Banking.
\end{enumerate}
A few examples of \rev{transformative} technologies leveraged in these areas include, \rev{the} use of AI for \rev{Asset Management} through bots as the new user interfaces, \rev{the} adoption of algorithmic trading that changed the stock market with High Frequency Trading for \rev{Investment Banking}, and the increased use of mobile devices for payments impacting financial transactions in \rev{Retail Banking, see also Figure \ref{fig:TransformativeTech}}.

\begin{figure}[htbp]
	\centering
		\includegraphics[width=.45\textwidth]{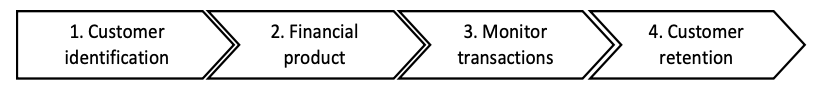}
	\caption{Customer life cycle conceptual design.
	\rev{Here, a customer can be a corporation, fund, financial institution, government, or individual.}}
	\label{fig:benefits-pipeline}
\end{figure}

\rev{The} adaption of new technologies and evolution of financial services can be traced in waves leading to a digital future. \rev{The} first wave focused on mobile adoption to engage customers. This was followed by the exploration of new business models to leverage new data sources and the threat from technology providers. The third, current wave is driven by \rev{a} combination of challenges including cost pressures, technology disruptions and regulatory changes\rev{, see Fig.~\ref{fig:TransformationWaves}. This is} causing a fundamental redefinition of the financial institution, \rev{its} internal operations\rev{,} and how it engages externally.
\rev{Further background is provided in Ref.~\cite{Bhaskaran2019}.}
\rev{Within this context, the future} benefits of quantum computing for businesses \rev{could} be measured across a set of key business metrics:
\begin{itemize}
\item
Reduce regulatory penalty costs or avoidable human labor.
\item
Improve customer satisfaction and brand perception.   
\item
Increase customer interaction and financial activity.
\item
Reduce capital levels and improve cash-flow.
\end{itemize}
\rev{In the following sections, we group specific problems arising in the financial services where classical computers face challenges or are insufficient, in three classes, \textbf{Simulation}, \textbf{Optimization} and \textbf{Machine learning}, we introduce the quantum algorithms applicable to them, and discuss results obtained on IBM Quantum back-ends for some specific problems.
Each quantum algorithm is applied on tasks and calculations that affect one or more phases of the customer life cycle, shown in Fig.~\ref{fig:benefits-pipeline}, and we describe the specific business benefits it may bring.} 

\section{Simulation}
\label{sec:simulation}

In this section we discuss simulation problems, where quantum computing may be beneficial. There are simulation problems at each stage of the customer life cycle (see Figure \ref{fig:benefits-pipeline}):

\begin{enumerate}
\item
\textbf{Customer identification}
Obtain new revenue sources for value-added services such as derivative pricing using sophisticated quantum computing algorithms. This offering can help compensate for the monetary losses of MiFIDII new transparency measures in trading, estimated to cost \$ 240 Million \cite{mifid2017}.
\item
\textbf{Financial products}
Better manage Value at Risk and the Economic Capital Requirement providing more accurate estimates to improve liquidity management by actively managing the balance sheet, increase capital to maintain a 7\% equity capital ratio to its riskier assets and avoid Basel III  related compliance fines \cite{reuters2011} that are up to 10\% of Bank's turnover (revenue).
\item
\textbf{Monitor transactions}
Allow for a more precise approach to incorporating market volatility into an institution's Tier 1 reporting \cite{VOB}, optimizing risk weighted assets results through a much more accurate/precise calculation process.  
\item
\textbf{Customer retention}
Improve risk analysis for the new Net Stable Funding Ratio time-frames requirements that will impact the cost of doing business for Prime Brokers and hedge funds. The stable funding for securities Lending Transactions shifts from 0\% to 10\% for Level 1 collateral and to 15\% for other collateral \cite{nomura2014}. \end{enumerate}

Overall, Basel regulation implementation and the new needs of risk management is of mandatory interest for financial institutions, with an estimated technology cost between EUR 45 million and EUR 70 Million \cite{mckinsey2010}.\\

Simulation focuses on creating scenarios of potential outcomes, such as the impact of volatility on risk, evaluating asset values for pricing, or monitoring economic system impacts in the market.

One key task in the finance industry \rev{where} simulation is crucial is the pricing of financial instruments and estimating their risk.
For example, the buyers and sellers of complex financial instruments gain a competitive advantage when they can price such instruments with a better accuracy than their competition.
Regulations such as Basel III require banks to perform stress tests and to hold an amount of capital that depends on their risk-weighted assets.
However, pricing and estimating the risk of many financial instruments is computationally intensive.
Analytical models are often too simplistic to capture the complex dependencies between financial instruments or cannot take into account some of their features such as path-dependency.
Monte Carlo (MC) simulations are, therefore, used to estimate risk metrics, to price financial instruments, and to perform scenario analysis that can be used in stress tests.
In a MC simulation $M$ samples are drawn from the model input distributions and are used to construct an estimation of a quantity of interest.
The confidence interval of this estimation scales as $\mathcal{O}(1/\sqrt{M})$ making MC computationally intensive. 
For instance, decreasing the size of the confidence interval by an order of magnitude requires increasing the computational cost by a factor of 100.
In this section, we briefly discuss MC for option pricing (see Sec.~\ref{sec:opt_price}) and risk calculations (see Sec.~\ref{sec:risk}) and then discuss how such tasks can be performed on quantum computers using amplitude estimation (see Sec.~\ref{sec:ae}).

\paragraph{Option Pricing\label{sec:opt_price}}
Options are financial derivative contracts that give the buyer the right, but not the obligation, to buy (call option) or sell (put option) an underlying asset at an agreed-upon price (strike) and time-frame (exercise window).
In their simplest form, the strike price is a fixed value and the time-frame is a single point in time, but exotic variants may be defined on more than one underlying asset, the strike price can be a function of several market parameters and could allow for multiple exercise dates \cite{Hull2006}.
Options provide investors with a vehicle to profit by taking a view on the market or exploit arbitrage opportunities and are core to various hedging strategies. 
As such, understanding their properties is a fundamental objective of financial engineering. 
Due to the stochastic nature of the parameters options are defined on, calculating their fair value can be an arduous task.
Analytical models exist for the simplest types of options \cite{Black1973} but the simplifying assumptions on the market dynamics required for the models to provide closed-form solutions often limit their applicability \cite{Dupire1994}.
Hence, more often than not, numerical MC simulations are employed for option pricing  since they are flexible and can generically handle stochastic parameters \cite{Boyle1977, Glasserman2003}.
Option pricing with MC generally proceeds by simulating many paths of the time evolution undergone by the underlying assets to build a distribution of the option payoff at maturity.
The option price is then obtained by discounting the expected value of this distribution.
Classical MC methods require extensive computational resources to provide accurate option price estimates, particularly for complex options.
Because of the widespread use of options in the finance industry, accelerating the methodology to price them can significantly impact the operations of a financial institution.

\paragraph{Risk Management\label{sec:risk}}
Risk management plays a central role in the financial system.
It allows companies, institutions, and individuals to avoid monetary losses and grow their business.
Financial risk, which comes in many forms such as credit risk, liquidity risk, and market risk, is often estimated using models and simulations.
The accuracy of these models has a direct impact on the operations of the entity using them.
For instance, the capital requirements imposed on banks under the Basel accords depend on the accuracy of risk models \cite{BaselIII2019}.
Therefore, banks with more accurate models can make better use of their capital. 
Value at risk (VaR) \cite{Glasserman2000}, a quantile of the loss distribution, is a widely used risk metric.
For example, the Basel III regulations require banks to perform stress tests using VaR \cite{BaselIII}. 
Monte Carlo simulations are the method of choice to determine VaR and CVaR.
They are done by building a model and computing the loss/profit distribution for $M$
different realizations of the model input parameters.
Many different runs are needed to achieve a representative distribution of the loss/profit distribution.
Classical attempts to improve the performance are variance reduction or Quasi-Monte Carlo techniques \cite{Corwin1996, Glasserman2003, Glasserman2005}.
The first aims at reducing the constants while not changing the asymptotic scaling; whereas, the latter improves the asymptotic behavior, but only works well for low-dimensional problems.

In Section \ref{sec:ae} we discuss how Quantum Amplitude Estimation can provide a quadratic speed-up over classical Monte Carlo (MC) simulations and highlight the steps needed to calculate value at risk in Section \ref{sec:var_ae}.
We then employ these methods in the context of credit risk as already discussed in \cite{Egger2019} and summarized here in Section \ref{sec:credit_risk}. For a detailed discussion on how Quantum Amplitude Estimation can provide an advantage for options pricing, we refer to \cite{Rebentrost_2018, Stamatopoulos_2020, Carrera2020}.

\subsection{Quantum Amplitude Estimation}\label{sec:ae}
\begin{figure}[t!]
	\centering
		\includegraphics[width=.45\textwidth]{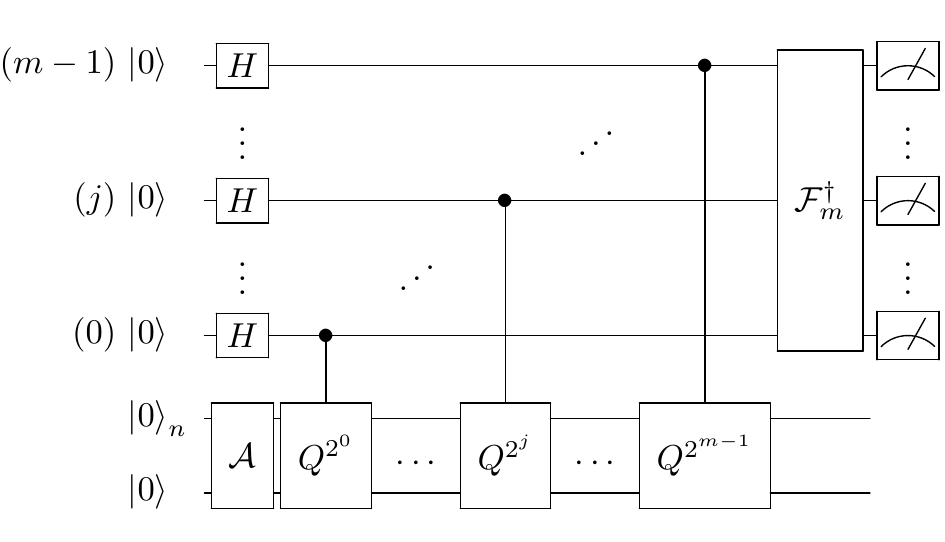}
	\caption{Quantum circuit for amplitude estimation as introduced in \cite{Brassard2000}. $H$ is the Hadamard gate and $\mathcal{F}_m^{\dagger}$ denotes the inverse Quantum Fourier Transform on $m$ qubits.}
	\label{fig:amplitude_estimation_circuit}
\end{figure}

Amplitude estimation (AE) is a quantum algorithm that can estimate a parameter $a$ with a convergence rate $\mathcal{O}(1/M)$ where $M$ is the number of quantum samples.
This corresponds to a quadratic speed-up compared to classical MC.
AE is based on a unitary operator $\mathcal{A}$ acting on a register of $(n+1)$ qubits such that $\mathcal{A} \ket{0}_{n+1} = \sqrt{1 - a}\ket{\psi_0}_n\ket{0} + \sqrt{a}\ket{\psi_1}_n\ket{1}$ for some normalized states $\ket{\psi_0}_n$ and $\ket{\psi_1}_n$, where $a \in [0, 1]$ is unknown.
AE allows the efficient estimation of $a$, i.e., the probability of measuring $\ket{1}$ in the last qubit \cite{Brassard2000}. 
This is done using an operator $Q$ defined as
\begin{equation*}
Q \!=\! \mathcal{A}(\mathbb{I} - 2 \ket{0}_{n+1} \bra{0}_{n+1})\mathcal{A}^{\dagger} (\mathbb{I} - 2 \ket{\psi_0}_{n}\ket{0} \bra{\psi_0}_{n}\bra{0}),
\end{equation*}
where $\mathbb{I}$ denotes the identity operator, 
and Quantum Phase Estimation \cite{Kitaev1995} to approximate certain eigenvalues of $Q$.
AE requires $m$ additional qubits and $M = 2^m$ (controlled) applications of $Q$.
The $m$ qubits are first put into equal superposition\rev{, i.e. $\ket{0}_m\to 1/\sqrt{M}\sum_{k=0}^{M-1}\ket{k}_m$, by applying Hadamard gates which are single-qubit gates that perform $\ket{0}\to (\ket{0}+\ket{1})/\sqrt{2}$.
Then, the $m$ qubits are used to control the applied power of $Q$, which leads to the state
\begin{eqnarray}
1/\sqrt{M} \sum_{k=0}^{M-1} \ket{k}_m Q^k \mathcal{A} \ket{0}_{n+1},
\end{eqnarray}
where each $Q^k$ imparts the phase $e^{-2ik\theta_a}$ and $e^{+2ik\theta_a}$, respectively, to the two eigenstates of $\mathcal{Q}$ that lie in the span of the states $\ket{\psi_0}_n\ket{0}$ and $\ket{\psi_1}_n\ket{1}$.
Next, the inverse Quantum Fourier Transform 
\begin{align}
\mathcal{F}_M^\dagger:\ket{k}_m\to\frac{1}{\sqrt{M}}\sum_{y=0}^{M-1}e^{-2\pi i yk/M}\ket{y}_m
\end{align}
is applied to interfere these phases.
Last, measuring the $m$ qubits results in an integer $y \in \{0, \ldots, M-1\}$ which corresponds to the states for which the phases interfered constructively, i.e., $y\simeq M\theta_a/\pi$, see the circuit in Fig. \ref{fig:amplitude_estimation_circuit}.
The integer $y$ is then} classically mapped to the estimator $\tilde{a} = \sin^2(y\pi/M) \in [0, 1]$.
The estimator $\tilde{a}$ satisfies
\begin{eqnarray}
| a - \tilde{a} | 
&\leq& \frac{2\sqrt{a(1-a)}\pi}{M} + \frac{\pi^2}{M^2}
\nonumber \\
&\leq& \frac{\pi}{M} + \frac{\pi^2}{M^2} \;=\; O\left(M^{-1}\right),\label{eq:ae_error_bound}
\end{eqnarray}
with probability of at least $8/\pi^2$.
This represents a quadratic speedup compared to the $O\left(M^{-1/2}\right)$ convergence rate of classical Monte Carlo methods \cite{Glasserman2000}.

Recently, several variants of AE have been proposed that simplify the required quantum circuits \cite{Suzuki2019, aaronson2019quantum, Grinko2019}.
They leverage the same underlying structure, but allow to remove the $m$ additional qubits as well as the Quantum Fourier Transform.
For a comparison of available AE variants, we refer to \cite{Grinko2019}.

\subsection{Estimating value at risk with AE} \label{sec:var_ae}
The discussion above shows that to efficiently estimate a parameter $a$ we need the corresponding operator $\mathcal{A}$.
We make use of AE in finance by building the $\mathcal{A}$ operator for each quantity of interest, such as a risk measure, of a random variate $X$.
We represent the distribution of $X$ as an $n$-qubit quantum state
\begin{align}
    \ket{\psi}_n=\sum_{i=0}^{N-1}\sqrt{p_i}\ket{i}_n
\end{align}
by discretizing the outcomes of $X$ and mapping them to $i\in\{0,...,N-1\}$ where $N=2^n$.
Here, $p_i\in[0,1]$ is the probability of measuring the state $\ket{i}_n$ which is a binary representation of $i$.
By adding an ancilla qubit to the $n$-qubit register and applying the operator
\begin{align}
    F:\ket{i}_n\ket{0}\to\ket{i}_n\left(\sqrt{1-f(i)}\ket{0}+\sqrt{f(i)}\ket{1}\right),
\end{align}
for some function $f(i)$, to the state $\ket{\psi}_n\ket{0}$ results in the quantum state
\begin{align}
    \sum_{i=0}^{N-1}\sqrt{p_i}\sqrt{1-f(i)}\ket{i}_n\ket{0}+\sum_{i=0}^{N-1}\sqrt{p_i}\sqrt{f(i)}\ket{i}_n\ket{1}.
\end{align}
By comparing this state to $\mathcal{A} \ket{0}_{n+1} = \sqrt{1 - a}\ket{\psi_0}_n\ket{0} + \sqrt{a}\ket{\psi_1}_n\ket{1}$ we find that the probability of measuring $\ket{1}$ in the ancilla qubit is $a=\sum_{i=0}^{N-1}p_if_i$.
We can therefore obtain an estimate for ${\rm VaR}_\alpha(X)$ by choosing
\begin{align}
    f(i)=\left\{ 
    \begin{array}{cc}
        1 & i\leq l \\
        0 & \text{otherwise}
    \end{array}
    \right.
\end{align}
for some level $l$.
With this definition of $f(i)$ the probability of measuring $\ket{1}$ in the ancilla qubit is $\sum_{i=0}^{l}p_i=\mathbb{P}[X\leq l]$.
With a binary search over $l$ we find the smallest $l_\alpha$ such that $\mathbb{P}[X\leq l_\alpha]\geq 1-\alpha$.
The smallest $l_\alpha$ corresponds to the value at risk.
This estimation of ${\rm VaR}_\alpha(X)$ has accuracy $\mathcal{O}(M^{-1})$, i.e. a quadratic speed-up compared
to classical Monte Carlo methods (omitting the additional
logarithmic complexity of the bisection search) -- \rev{where $\mathcal{O}(\cdot)$ indicates the big-$\mathcal{O}$ notation}.
\rev{Here, we assumed a one-dimensional probability distribution.
Probability distributions with more then one random variable can be loaded into a corresponding number of qubit registers \cite{Woerner2019}.
Correlations between the random variables are then introduced by suitably entangling these qubit registers.
}

\subsection{Credit risk}\label{sec:credit_risk}
The quantum method to calculate value at risk, outlined in the previous sections, can be applied in the context of credit risk to determine the economic capital requirement (ECR) associated to holding a portfolio of $K$ loans \cite{Egger2019}.
The ECR is the amount of capital that needs to be held on the balance sheet to protect against unexpected losses. It is therefore defined as the value at risk less the expected value of the loss distribution $\mathcal{L}$, i.e.
\begin{align}
    {\rm ECR}_\alpha(\mathcal{L}) = {\rm VaR}_\alpha(\mathcal{L})-\mathds{E}(\mathcal{L})
\end{align}
where $\alpha$ is the confidence level. Estimating the value at risk is often a computationally intensive task requiring classical MC simulation. However, quantum amplitude estimation can achieve the same result with a quadratic speed-up.
We illustrate amplitude estimation for a portfolio of $K$ assets for which the multivariate random variable $(L_1, ..., L_K) \in \mathbb{R}_{\geq 0}^K$ denotes each possible loss associated to each asset.
The expected value of the total loss $\mathcal{L} = \sum_{k=1}^K L_k$ is $\mathbb{E}[\mathcal{L}] = \sum_{k=1}^{K} \mathbb{E}[L_k]$.
The value at risk for a given confidence level $\alpha \in [0, 1]$ is defined as the smallest total loss that still has a probability greater than or equal to $\alpha$, i.e.,
\begin{eqnarray}
\text{VaR}_{\alpha}[\mathcal{L}] &=& \inf_{x \geq 0} \left\{ x \mid \mathbb{P}[\mathcal{L} \leq x] \geq \alpha \right\}.
\end{eqnarray}
Common values of $\alpha$ for ECR found in the finance industry are around $99.9\%$.
We assign a Bernoulli random variable $X_k$ to each asset to indicate if it is in a default state such that $L_k = \lambda_k X_k$ where $\lambda_k > 0$ is the loss given default (LGD).
The probability that $X_k=1$, i.e., a loss for asset $k$, is $p_k$.
The expected loss of the portfolio $\mathbb{E}[\mathcal{L}] = \sum_{k=1}^K \lambda_k p_k$ is easier to evaluate than $\text{VaR}_{\alpha}[\mathcal{L}]$, which usually requires a Monte Carlo simulation.
The defaults $X_k$ are usually correlated which we model following a conditional independence scheme \cite{Rutkowski2014}.
Given a realization $z$ of a latent random variable $\mathcal{Z}$, the Bernoulli random variables $X_k \mid \mathcal{Z}=z$ are assumed independent, but their default probabilities $p_k$ depend on $z$.
We follow \cite{Rutkowski2014} and assume that $\mathcal{Z}$ has a standard normal distribution and that
\begin{eqnarray}
p_{k}(z) &=& \Phi\left( \frac{\Phi^{-1}(p_k^0) - \sqrt{\rho_k} z}{\sqrt{1 - \rho_k}} \right),
\end{eqnarray}
where $p_k^0$ denotes the default probability for $z=0$, $\Phi$ is the cumulative distribution function (CDF) of the standard normal distribution, and $\rho_k \in [0, 1)$ determines the sensitivity of $X_k$ to $\mathcal{Z}$.
This scheme is similar to the one used for regulatory purposes in the Basel II (and following) Internal Ratings-Based (IRB) approach to credit risk \cite{BaselII, BaselIII}, and is called the \emph{Gaussian conditional independence model} \cite{Rutkowski2014}. 

To estimate VaR, we use AE to efficiently evaluate the CDF of the total loss, i.e., we will construct $\mathcal{A}$ such that  $a = \mathbb{P}[\mathcal{L} \leq x]$ for a given $x \geq 0$, and apply a bisection search to find the smallest $x_{\alpha} \geq 0$ such that $\mathbb{P}[\mathcal{L} \leq x_{\alpha}] \geq \alpha$, which implies $x_{\alpha} = \text{VaR}_{\alpha}[\mathcal{L}]$ \cite{Woerner2019}.

\begin{figure}[hbtp]
\centering
\includegraphics[width=0.475\textwidth]{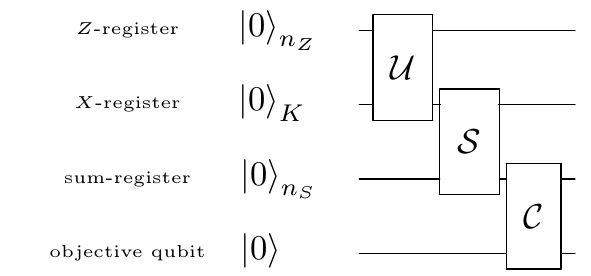}
\caption{\label{fig:high_level_cdf_circuit} High level circuit of the operator $\mathcal{A}$ used to evaluate the CDF of the total loss: the first qubit register with $n_Z$ qubits represents $\mathcal{Z}$, the second qubit register with $K$ qubits represents the $X_k$, the third qubit register with $n_S$ qubits represents the sum of the losses, i.e., the total loss, and the last qubit is flipped to $\ket{1}$ if the total loss is less than or equal to a given $x$.
The operators $\mathcal{U}$, $\mathcal{S}$, and $\mathcal{C}$ represent the loading of uncertainty, the summation of losses, and the comparison to a given $x$, respectively.}
\end{figure}

Mapping the CDF of the total loss to a quantum operator $\mathcal{A}$ requires three steps. Each step corresponds to a quantum operator.
First, $\mathcal{U}$ loads the uncertainty model. Second, $\mathcal{S}$ computes the total loss into a quantum register with $n_S$ qubits. Last, $\mathcal{C}$ flips a target qubit if the total loss is less than or equal to a given level $x$ which is used to search for $\text{VaR}_\alpha$.
Thus, we have $\mathcal{A} = \mathcal{CSU}$ and the corresponding circuit is illustrated in Fig.~\ref{fig:high_level_cdf_circuit} on a high level.

We now discuss the operators $\mathcal{U}$, $\mathcal{S}$, and $\mathcal{C}$ in more detail.
The loading operator $\mathcal{U}$ loads the distribution of $\mathcal{Z}$ and prepares the $X_k$ of each asset.
To include correlations between the default events we represent $\mathcal{Z}$ in a register with $n_Z$ qubits.
We use a truncated and discretized approximation with $2^{n_Z}$ values, where we consider an affine mapping $z_i = a_z i + b_z$ from $i \in \{0, ..., 2^{n_Z}-1\}$ to the desired range of values of $\mathcal{Z}$. 
Since $\mathcal{Z}$ follows a standard normal distribution we can efficiently load it to a quantum register with controlled rotations \cite{grover2002logconcave}.
We encode the $X_k$ of each asset in the state of a corresponding qubit by applying to qubit $k$ a $Y$-rotation $R_Y(\theta_p^k)$, controlled by the qubit register representing $\mathcal{Z}$, with angle $\theta_p^k(z) = 2 \arcsin(\sqrt{p_{k}(z)})$. 
For simplicity, we use a first order approximation\footnote{Higher order approximations of $\theta_p^k(z)$ can be implemented using multi-controlled rotations.
Furthermore, by using quantum arithmetic one could also compute $\theta_p^k(z)$ directly \cite{Woerner2019}.} of $\theta_p^k(z)$ and include the affine mapping from $z$ (a value of the normal distribution) to $i$ (an integer represented by $n_Z$ qubits), i.e., $\theta_p^k(z_i) \approx a_k i + b_k$.
This prepares qubit $k$ in the state $\sqrt{1 - p_k}\ket{0} + \sqrt{p_k}\ket{1}$ for which the probability to measure $\ket{1}$ is $p_k$. 
The $\ket{1}$ state of qubit $k$ thus corresponds to a loss for asset $k$.

Next, we need to compute the resulting total loss for every realization of the $X_k$.
Therefore, we use a weighted sum operator
\begin{eqnarray}
&\mathcal{S}: & \ket{x_1, \cdots, x_K}_K \ket{0}_{n_S} \nonumber \\ 
&\mapsto & \ket{x_1, \cdots, x_K}_K \ket{ \lambda_1 x_1 + \cdots + \lambda_K x_K}_{n_S},
\end{eqnarray}
where $x_k \in \{0, 1\}$ denote the possible realizations of $X_k$.
We set the size of the sum register to $n_S = \lfloor\log_2(\lambda_1 + \cdots + \lambda_K)\rfloor + 1$ qubits to represent all possible values of the sum of the losses given default $\lambda_k$, assumed to be integers.
To implement $\mathcal{S}$ we apply a divide and conquer approach and first sum up pairs of assets, then pairs of the resulting sums and so on until we computed the total sum.
This implies that we start with a weighted-sum operator, discussed in detail in \cite{Stamatopoulos_2020}, and then continue with adder circuits \cite{Cuccaro2004} to iteratively combine the intermediate results.

Last, we need an operator that compares a particular loss realization to a given $x$ and then flips a target qubit from $\ket{0}$ to $\ket{1}$ if the loss is less than or equal to $x$.
This operator is defined by
\begin{eqnarray}
\mathcal{C}: \ket{i}_{n_S}\ket{0} \mapsto 
\begin{cases}
\ket{i}_{n_S}\ket{1} & \text{if $i \leq x$}, \\
\ket{i}_{n_S}\ket{0} & \text{otherwise.}
\end{cases}
\end{eqnarray}
A fixed value comparator, i.e., a comparator that takes a fixed value to compare to as a classical input, can be based on adder circuits \cite{Draper2004}.
Here, a logarithmic scaling can be achieved by adding a linear number of ancilla qubits.

\begin{figure}
    \centering
    \includegraphics[width=0.4\textwidth]{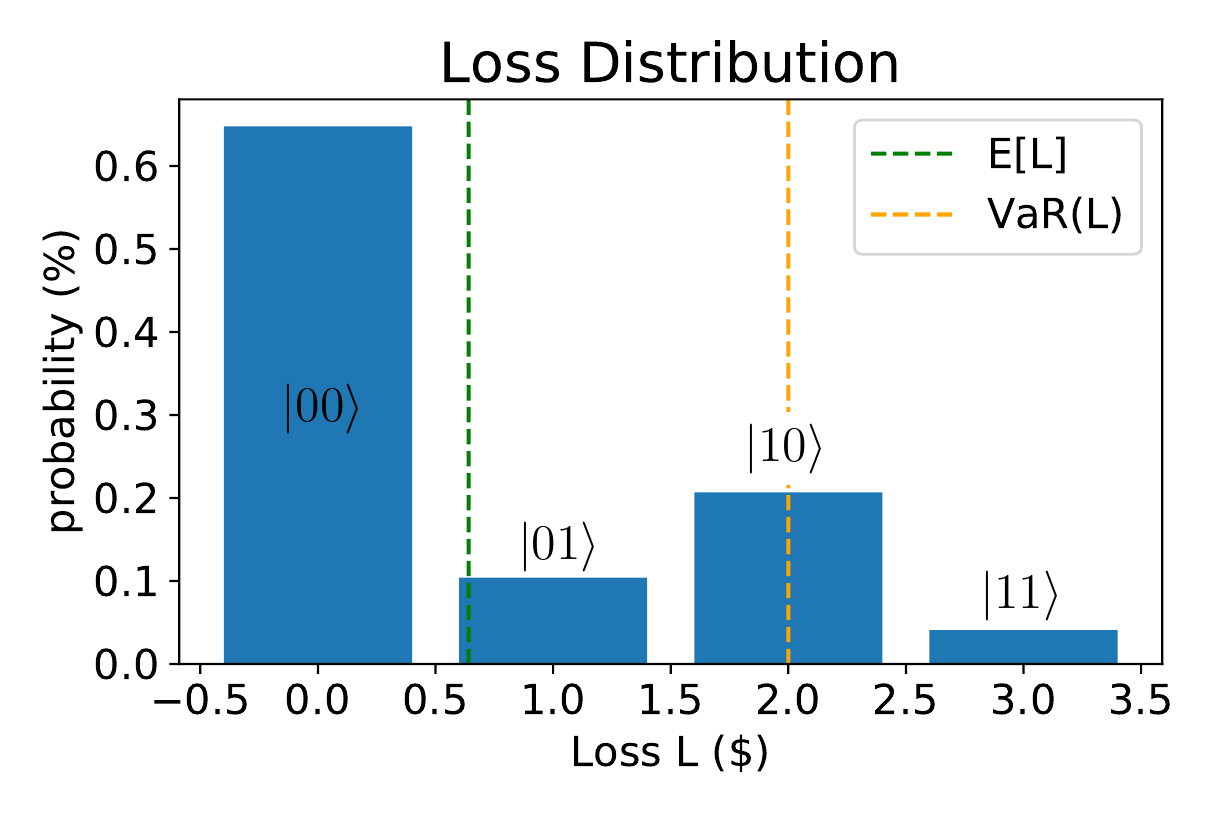}
    \caption{Loss distribution of the two-asset portfolio. The green and orange dashed lines show the expected value and the 95\% value at risk of the distribution, respectively.}
    \label{fig:loss}
\end{figure}

We now show the performance of the quantum algorithm for an illustrative example with $K=2$ assets.
Each asset is defined by the triplet $(\lambda_k, p_k^0, \rho_k)$, i.e. the loss given default, the default probability for $z=0$, and the sensitivity of $X_k$ to $\mathcal{Z}$, respectively.
We chose $(1,\, 0.15,\, 0.1)$ and $(2,\, 0.25,\, 0.05)$ for asset one and two, respectively, which results in the loss distribution shown in Fig.~\ref{fig:loss}.
\rev{This distribution is naturally encoded into the four states of two qubits, indicated by the state labels in Fig.~\ref{fig:loss} which illustrates how a qubit register can encode the exponential number of portfolio states as a quantum superposition.}
From the chosen $\lambda_k$'s it follows that the sum register requires $n_S = 2$ qubits to represent all possible losses.
We represent $\mathcal{Z}$ with $n_Z=2$ qubits.
Thus, $\mathcal{A}$ is operating on seven qubits that represent this problem on a quantum computer, including the objective qubit.

\begin{figure}
    \centering
    \includegraphics[width=0.48\textwidth]{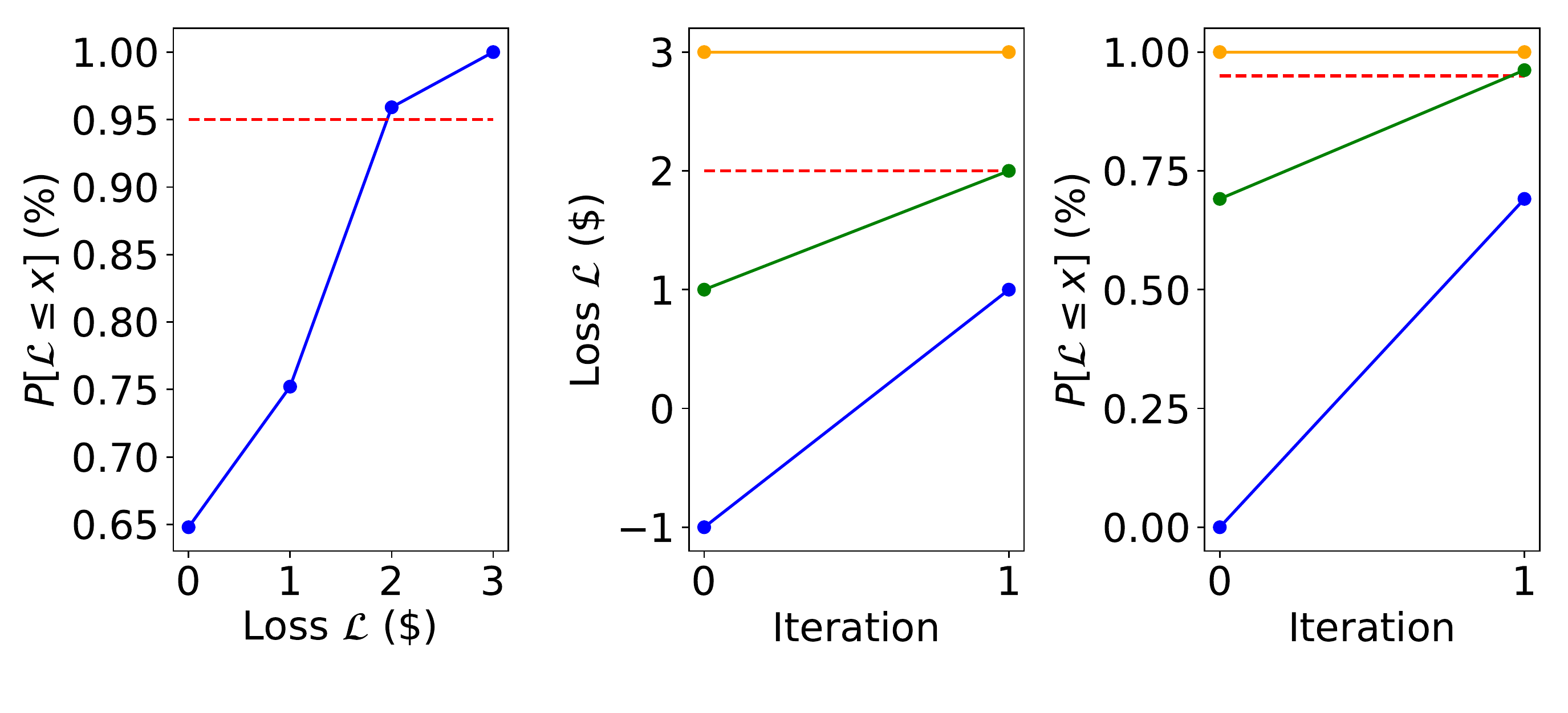}
    \caption{Cumulative distribution function (left) of the total loss $\mathcal{L}$ (blue), shown in Fig.~\ref{fig:loss}, for a two-asset portfolio. The red dashed line is the target value at risk level $\alpha=95\%$.
    Bisection search to compute VaR (middle and right) using $m=4$ sampling qubits, i.e. 16 quantum samples: orange and blue represent the lower and upper bounds of the search interval, respectively, and green is the resulting midpoint. The red dashed line shows the exact value.}
    \label{fig:var_bisection_search}
\end{figure}

To simulate our algorithm we input the circuit for $\mathcal{A}$ to the AE sub-routine implemented in \emph{Qiskit}~\cite{Qiskit} and perform the bisection search using the result to find $x_\alpha$.
We use $m=4$ evaluation qubits giving us 16 quantum samples.
Our implementation requires one additional ancilla qubit to create $\mathcal{Q}$.
Therefore, this experiment  requires a total of 12 qubits that we simulate on classical computers using the \texttt{statevector\_simulator} back-end provided by \emph{Qiskit Aer}.
Since $n_S=2$, the bisection search requires at most two steps, as shown in Fig.~\ref{fig:var_bisection_search}.
To ensure that the entire probability distribution is captured by the initial lower and upper bounds of the bisection search we set them at losses of -1\$ and 3\$, respectively.
The simulations, shown in Fig.~\ref{fig:var_bisection_search}, properly identify the 95\% value at risk, located at a loss of 2\$, on the first iteration of the bisection search.
\rev{We expect that the number of iterations needed in the bisection search will scale linearly with the number of assets in the portfolio.}

\rev{We have investigated in Ref.~\cite{Egger2019} the quantum resources required for $K=2^{20}$ assets, i.e. a portfolio of approximately one million assets.
With $n_Z=10$, $n_S=30$, and $m=10$, a fault tolerant universal quantum computer would require approximately 37 million T/Toffoli gates.
If we assume that error-corrected T/Toffoli-gates can be executed in $10^{-4}$ seconds \cite{Fowler2018} and that the quantum phase estimation can be removed \cite{Suzuki2019}, which reduces the circuit depth by a factor of two, we estimate a runtime of 30 minutes to estimate the value at risk for a one-million-asset portfolio \cite{Egger2019}.
This estimate may change as quantum computing technology advances.}

\subsection{Distribution loading}
Replacing a MC simulation with \rev{AE} requires efficiently loading the distributions of the random variables in the model to the quantum computer to avoid diminishing the potential Quantum Advantage.
This is feasible, e.g. for efficiently integrable probability distributions such as log-concave distributions for which the loading operator can be built from controlled rotations \cite{grover2002logconcave}.
The loading of arbitrary states into quantum systems, however, requires exponentially many gates \cite{Plesch2011}, making it inefficient to model arbitrary distributions as quantum gates.
Since the distributions of interest are often of a special form, the limitation may be overcome by using quantum Generative Adverserial Networks (qGAN). 
These networks allow us to load a distribution using a polynomial number of gates \cite{zoufal2019qgan}.

\subsection{Summary}

We have shown how AE can be used to estimate the ECR for a portfolio of loans.
This results in a quadratic speed-up (omitting the logarithmic cost of the bisection search in VaR) over classical MC simulations.
The example also shows that the quantum circuit needed to implement $\mathcal{A}$ depends on the task at hand.
Therefore, extending this work to other financial simulation tasks requires task specific quantum circuits to implement $\mathcal{A}$.

\section{Optimization}
\label{sec:optimization}

In this section we discuss optimization problems, where quantum computing may be beneficial. As in the case of simulation, there are optimization problems at each stage of the customer life cycle (see Figure \ref{fig:benefits-pipeline}):

\begin{enumerate}
\item
\textbf{Customer identification (and assesment)}
Improve Financial Supply Chain Efficiency \cite{treasury2013}  in procurement and payment focusing on customers and suppliers to increase to reduce working capital levels, enhance liquidity, minimize risk and avoid late payments (47\% of suppliers are paid late \cite{FSCM2016}).
\item
\textbf{Financial products}
Accelerate trade settlement capacity \cite{DTCC2018, braine2019quantum}  (i.e. from 45\% transactions to 90\%) to reduce associated capital requirements, systemic risk, and operational costs.
\item
\textbf{Monitor transactions}
Keep investment portfolios relevant by re-balancing aligned to market changes \cite{CNBC2017}, while handling all the associated fees (taxes, commissions, etc.). This can reduce transaction costs by 50\% and lead to \$ 600k savings in trading costs for an example of 4 asset \$~1billion portfolio \cite{IPE}.
\item
\textbf{Customer retention}
 Improve the process of matching companies to potential buyers to avoid current customer churn towards automated investment banking. This can reduce current work losses, in 2015, 26\% of the \$1B+  merger and acquisitions were done without involvement of financial advisors \cite{CBInsights}. 
\end{enumerate}

Overall, investment banks are increasingly applying technology to automate the trading pipeline, hiring technologists, which are 20\% to 40\% job openings on the main investment banks \cite{GS2020}. Also, JPMorgan spends the most a year on technology with \$ 10.8 billion \cite{OutsideInsight}.

Many financial services firms may want to take actions that result in the best possible outcome for a given goal. 
In the language of mathematical optimization, finding the best decision or action with respect to maximizing or minimizing given goals or objectives, 
is cast as  maximizing or minimizing an objective function in a decision variable, subject to constraints, often given again by functions of a decision variable.
This has extensive applications, e.g., finding the best supply-chain route for delivery, determining the best investment strategy for a portfolio of assets, or increasing productivity with a number of fixed resources in operations.

Optimization problems in finance may consider a single period, 
where all information are available at time 0 and one takes a one-off decision,
or their generalizations.
In multi-stage problems, information become available at multiple points in time,
and also the decisions can be made at multiple points in time (stages).
In optimal control problems, one optimizes over policies, which drive the 
repeated decisions.
Throughout, one can work with either discrete decisions (e.g., yes/no, number of round lots), 
or real-valued decisions (e.g., price).
Throughout, one can enforce the constraints given by the Basel III regulatory framework directly,
or produce the decision that would be optimal without the constraints, and test whether these constraints are satisfied using simulation tools, as introduced above.

Correspondingly, there is a breadth of approaches, which model active and passive investment management. 
Within active investment management, one often tries to find the optimal investment strategy striking a balance between the expected profit and some measure of risk involved.
Within passive investment management, one can imagine index-tracking funds and their ``calibration problems'', which are based on portfolio diversification, and aim at representing a portfolio with a large number of assets by a smaller number of representative assets. 
Within auction mechanisms, the clearing of so-called combinatorial auctions, where bids on a subset of items are accepted, is an example of a difficult discrete-valued optimization problem.
In this section, we introduce idealized versions of these problems, 
where only the decision in the next period is considered, 
transaction costs are ignored, and 
Basel III constraints are not enforced, and the corresponding quantum algorithms.

\rev{As we will discuss shortly, different quantum algorithms have been advocated for different optimization problem classes, e.g., the algorithms for convex problems are different from the ones for discrete problems. } \rev{In discrete problems, e.g., we will discuss variational approaches, such as VQE and QAOA.} \rev{While for other applications quantum computing offers more clear-cut benefits with respect to classical computing (e.g., Shor's algorithm), in the application of optimization algorithms these benefits are still an active research area, see for instance \cite{gambella2020multi,Brandao2018fixed,Huang2019, Shaydulin2019b,Guerreschi2019,nannicini2019performance,Zhou2018,Crooks2018,Gilliam2019,Brandao2019,Preskill2018,AaronsonBook}.} 
\rev{What it is fair to say is that with the advance of technology, quantum computing will play a major role in optimization and, in some problem classes, one will see tangible benefits, either in terms of solution quality, computational time, or both. }

\begin{table*}[t!]
\caption{An overview of recently proposed quantum algorithms for convex optimization. See Section~\ref{sec:cvxopt} for the notation, but note that $\tilde O$ hides polylogarithmic
terms in the upper bound.}
\sf
\label{tab:cvxopt}
\begin{tabularx}{520pt}{llX}
\toprule
Ref. & Quantum run-time & Comment \\ \midrule
\cite{8104077} & $\tilde  O(\sqrt{mn}s^2 \mathsf{P}^{32} /\delta^{18})$ & to error $1 \pm \delta$, where $\mathsf{P}$ is linear in $n$ in general \\
\cite{Brandao2018} & $\tilde  O(s^2 \textrm{poly}(\mathsf{P}\mathsf{D}/\epsilon) (\sqrt{m}/\epsilon^{10} +\sqrt{n}/\epsilon^{12} )$ \\
\cite{van2020quantum} &  $\tilde O(\sqrt{mn}s^2 (\mathsf{P}\mathsf{D}/\epsilon)^8)$ & where $\mathsf{P}$ is linear in $n$ in general \\
\cite{van2019improvements} &  $\tilde O((\sqrt{mn}s(\mathsf{P}\mathsf{D}/\epsilon)^{4})$ & where $\mathsf{P}$ is linear in $n$ in general, and quantum-read/classical-write \\
& & RAM can be accessed in constant time\\
\cite{kerenidis2018quantum} &  $\tilde O(n^{2.5} \mu \kappa^3 \log{1/\epsilon})$ &  where $\kappa$ goes to infinity for all instances \\
\bottomrule
\end{tabularx}
\end{table*}

\subsection{Problem classes: convex problems}
\label{sec:cvxopt}

First, we consider convex optimization, which encompasses linear programming (LP), quadratic programming (QP), and semidefinite programming (SDP). Convex optimization~\cite{boyd2004convex} is a sub-class of continuous optimization problems, where the decision variables are continuous, and it has been advocated for a large variety of applications, \rev{among which finance problems~\cite{Cornuejols2006}.}  

Not surprisingly, much of the recent interest in quantum algorithms for continuous optimization has focused on 
approaches to solving convex optimization problems, and in particular semidefinite programming (SDP). A SDP can be mathematically modelled as:

\begin{align}
\operatorname{inf}\ \langle C,X \rangle  
\textrm{ s.t. }  \mathcal{A}X=b, 
X\succeq_{\mathcal{K}}0 
\end{align}
where cone $\mathcal{K}$ is the cone of positive semidefinite symmetric $n\times n$ matrices  $\mathcal{S}_{+}^{n}$, i.e., $\{X=X^{\intercal}\in\mathbb{R}^{n \times n} |\ X \text{ is positive semidefinite}\}$, and $\mathcal{A} \colon \mathcal{S}^{n} \to \mathbb{R}^{m}$ is a linear operator between $\mathcal{S}_{+}^{n}$ and $\mathbb{R}^{m}$:
\begin{align*}
  X &\mapsto \begin{pmatrix}
  \langle A_{1},X\rangle\\
  \dots\\
  \langle A_{m},X\rangle
  \end{pmatrix}
\end{align*}
This is a proper generalization of linear programming (LP), second-order cone programming (SOCP), and convex cases of quadratically-constrained quadratic programming (QCQP), and hence of considerable practical interest. 

Within classical algorithms, there exist polynomial-time algorithms that can solve SDP. In particular, there are classical upper bounds on the run time as 
$$
\mathcal{O}(m(m^2 + n\omega + m \nz) \textrm{polylog}(m, n, \mathsf{P}, 1/\epsilon)),
$$ 
where, \rev{as said}, $\mathcal{O}(\cdot)$ indicates the big-$\mathcal{O}$ notation, $n$ is the dimension of the problem, $\omega \in [2, 2.373)$ is the exponent for matrix multiplication, $m$ is the number of constraints, and  
$\nz$ is the maximal number of non-zero entries per row of the input matrices. $\mathsf{P}$ is an upper bound on the trace of an optimal primal solution of an SDP, which could be seen as bound on a diameter of a ball outscribed to feasible solutions, in a suitable norm. This bound shows that SDPs can be approximated to any $\epsilon$ in polynomial time classically.

As for quantum algorithms, the early papers of \cite{8104077} and \cite{van2020quantum} and \cite{van2019improvements}
quantized the so-called multiplicative-weight-update (MWU) algorithm
of Arora and Kale and variants by Hazan. As it has been shown in \cite[Appendix E]{van2020quantum}, in the MWU algorithm, $\frac{\mathsf{P} \mathsf{D}}{\epsilon}$
should be seen as an important parameter ($\mathsf{D}$ being the dual counterpart of $\mathsf{P}$), as one can trade-off dependence on one of the
three individual parameters for the dependence on the others.

Subsequently, the work in \cite{kerenidis2018quantum} attempted a translation of primal-dual interior-point methods to quantum computers, but due to their reliance on solving linear systems, ended up with a bound dependent on the condition number $\kappa$ of the Karush-Kuhn-Tucker (KKT) system, which goes to infinity for all instances, by the design of the method, which may be not ideal in practice.

Finally, in~\cite{van2020convex}, the authors study the relationship of several oracles useful in subgradient algorithms, but do not claim a run-time of a particular algorithm for SDPs.

The key results are summarized in Table~\ref{tab:cvxopt}. As it can be observed, some of the quantum algorithms listed report scaling 
with $O(\sqrt{mn})$ \cite{8104077} or even $O(\sqrt{m} \ \textrm{poly}(\log(m), \log n))$ \cite{Brandao2018}.
However, these these upper bounds hide the diameters of balls outscribed to the primal and dual solutions. That is: these upper bounds assume that parameters $\mathsf{P}$ and $\mathsf{D}$ are constants independent of dimension, which could be however hard to satisfy in practice. 

In fact, if ones assumes that $\mathsf{P}$ and $\mathsf{D}$ are dependent on the dimension of the problem, then lower bounds on the run-time of quantum algorithms can be derived, e.g., for continuous Markowitz portfolio optimization problems in 
Section~\ref{sec:Markowitz}. 

\rev{Before moving on to an actual finance application, it is useful to briefly outline the quantum part of a quantum SDP algorithm. We focus here on the one of~\cite{8104077}, since it seems to be the one that has spurred much of the following research. As said, the authors in~\cite{8104077}, quantize the classical MWU algorithm for solving SDPs. In particular, they replace two steps of the classical algorithm by quantum subroutines that are more efficient than classical ones. } \rev{Consequently, the quantum SDP algorithm is not purely quantum, which is also the case for VQE in combinatorial optimization.} \rev{The steps that are replaced are as follows. First, it turns out that one can use a  ``Gibbs sampler'' to prepare the new primal candidate as a  $log(n)$-qubit quantum state in much less time than needed to compute it as an $n \times n$ matrix. Second, one can efficiently implement an oracle that is needed in the algorithm based on a number of copies of the quantum state, and using those copies to estimate trace operations}. \rev{The resulting oracle is weaker than what is used classically, in the sense that it outputs a sample rather than the whole vector (as typical in quantum computing). However, such sampling still suffices to make the algorithm work}. \rev{The interested reader is referred to~\cite{van2020quantum, Brandao2019} for more technical details.}

\subsection{Modern Portfolio Management - Active Investment Management: continuous case}\label{sec:Markowitz}

Let us now consider modern portfolio management, and develop lower bounds on the run-time of any quantum algorithm in the quantum query model of Beals et
al. \cite{beals2001quantum},
where quantum computation with $T$ queries is a sequence interleaving 
 $T$ unitary and $T$ query (oracle) transformations, with a measurement at the end.
These lower bounds on the run-time are generally based on lower bounds on parameters $\mathsf{P}, \mathsf{D} = \Theta(\min\{n,m\}^2)$, as discussed in Section~\ref{sec:cvxopt}, and suggest that the quantum speed-up of MWU algorithms 
\cite{8104077,van2020quantum,van2019improvements}
may be limited in practice.

In modern portfolio theory, one often assumes that there are $n$ possible assets,
and a number $m$ of forecasts of their returns $b_i \in \mathbb{R}^n$, $1 \le i \le m$
based on some historical returns $c \in \mathbb{R}^n$,
with a known covariance matrix $\Sigma \in \mathbb{R}^{n \times n}$ of the returns.
Minimization of the risk subject to lower bounds $\mu_i$ on the forecast returns leads to:
\begin{align}
	\label{eq:Markowitz}
\min_{w \in \mathbb{R}_+^n}	w^{\intercal}\Sigma w 
\quad \textrm{ s.t. } b_i^{\intercal} w \ge \mu_i, \quad \forall\, 1 \le i \le m
\end{align}
possibly with normalization such as $\sum_{j = 1}^{n} w_i = 1$ or similar.
In the spirit of Markowitz, one may consider a linear combination of the returns and the risk: 
\begin{align}
	\label{eq:Markowitz2}
\max_{w \in \mathbb{R}_+^n}	c^{{\intercal}}w - q w^{\intercal}\Sigma w 
\quad \textrm{ s.t. } b_i^{\intercal}w \ge \mu_i, \quad \forall\, 1 \le i \le m.
\end{align}
One can also formulate a Lagrangian of the Markowitz model above
and maximize $c^{\intercal} w - q w^{\intercal}\Sigma w$,
where the higher $q \geq 0$, the more risk-averse the portfolio will be. 

In any case, due to the general result that there exist problem instances for which complexity of every quantum LP-solver (and hence also SDP-solver) is the same as classical~\cite{van2020quantum}, there are instances of Markowitz Portfolio  Management \eqref{eq:Markowitz2} with $n$ assets and $m$ forecasts of the returns, for which a quantum algorithm has the same complexity of a classical one. This has to be taken as an understanding that quantum algorithms may help in some instances of Markowitz Portfolio  Management but not in others, depending on the actual input data. 

\subsection{Problem classes: combinatorial problems}

We move now to overview combinatorial problems and quantum algorithms that have been advocated for them. Combinatorial optimization problems are the ones for which the decision variables can be also discrete. Combinatorial problems are in general non-convex and not solvable with polynomial-time algorithms, classically. In the quantum domain, variational algorithms for general mixed-binary constrained optimization problems have been studied, and we will overview them as well as apply them for financial problems \rev{\cite{moll2018quantum}}. Firstly, we will look at Quadratic Binary Unconstrained Optimization (QUBO), where VQE/QAOA heuristic approaches have been advocated on noisy quantum devices \rev{\cite{nannicini2019performance, farhi2014quantum}}. VQE stands for variational quantum eigensolver, while QAOA stands for quantum approximate optimization algorithm. Then, we will explore how the classical alternating direction method of multipliers (ADMM) can help solving certain classes of mixed-binary constrained optimization problems \rev{\cite{gambella2020multi}}. We note that, currently, there is no theoretical guarantee that
variational algorithms on quantum devices 
can achieve significant speed-ups for QUBOs and the algorithms reported in this section are used as heuristics. On the other hand, variational algorithms do have non-trivial provable guarantees and they are not efficiently simulatable by classical computers. They are thus appealing algorithms to explore on near-term quantum machines~\cite{Zhou2018}.

\subsection{Variational approaches for Quadratic Binary Unconstrained Optimization (QUBO)}\label{sec:alg-qubo}

The attempts to solve mathematical optimization problems on  early generation of universal quantum computers have mainly focused on variational approaches \cite{mcclean2016theory, farhi2017quantum, barkoutsos2019improving}. In broad terms, a variational approach works by choosing a parametrization of the space of quantum states that depends on a relatively small set of parameters, then using classical optimization routines to determine values of the parameters corresponding to a quantum state that maximizes or minimizes a given utility function. Typically, the utility function is given by a Hamiltonian encoding the total energy of the system, to be minimized. The variational theorem then ensures that the expectation value of the Hamiltonian is greater than or equal to the minimum eigenvalue of the Hamiltonian. A large problem class tackled by such variational approaches is that of quadratic unconstrained binary optimization problems (QUBO):
\begin{align*}
	\min_{x} && c^{\transp} x + x^{\transp} Q x &&\\
	\text{s.t.:}&& x \in \{0,1\}^n,&& 	\text{with $c \in \mathbb{R}^n, Q \in \mathbb{R}^{n \times n}$}
\end{align*}
Each QUBO can be transformed into an Ising model with Hamiltonian constituted as a summation of weighted tensor products of $Z$ Pauli operators, i.e., $Z = \begin{pmatrix}
    1 & 0\\ 0 & -1
	\end{pmatrix}$, by mapping the binary variables $x$ to spin variables $y \in \{-1, 1\}$, i.e., $x = \frac{y+1}{2}$. In case equality constraints $Ax=b$ are present in the mathematical programming formulation, a QUBO can still be devised by adding a quadratic penalization $\alpha \lVert Ax-b \rVert^2$ to the objective function, as a soft-constraint in an Augmented Lagrangian fashion \cite{fiacco1990nonlinear, wang2009unified, nannicini2019performance}.

A typical variational approach on quantum devices, such as VQE~\cite{peruzzo2014variational} would involve the following two key steps in solving a QUBO, given its Ising Hamiltonian $H \in \mathbb{C}^{n\times n}$. First, one would parametrize the quantum state via a small set of rotation parameters $\theta$: each state can then be expressed as $|\psi(\theta)\rangle = U(\theta)|0\rangle$, where $U(\theta)$ is the parametrized quantum circuit applied to the initial state $|0\rangle$. The variational approach would then aim at solving $\min_{\theta} \langle\psi(\theta)| H |\psi(\theta)\rangle$. Such optimization can be performed in a setting that uses a classical computer running an iterative algorithm to select $\theta$, and a quantum computer to compute information about $\langle\psi(\theta)| H |\psi(\theta)\rangle$ for given $\theta$ (e.g., its gradient). The algorithm outline of VQE is reported in Algorithm \ref{algo:VQE}.
	
\begin{algorithm}
	\caption{Outline of VQE}\label{algo:VQE} 
	\sf
	\begin{algorithmic}[1] 
	       \small{\REQUIRE Hamiltonian $H$. Set $\theta = \theta_0$
		   \WHILE{Error tolerance is unmet:}
		   \STATE{Quantum part:}
		   \begin{itemize}
          \item Form variational state $\ket{\psi({\theta})} = U(\theta)\ket{0}$
          \item Compute information about $\lambda = \langle\psi(\theta)| H |\psi(\theta)\rangle$
          \end{itemize}
          \STATE{Classical part:}
          \begin{itemize}
          \item Update $\theta$ via a classical optimization algorithm (e.g., COBYLA, SPSA, etc.)
          \item Compute the error metric
          \end{itemize}
          \ENDWHILE
          \RETURN $\lambda, \theta$}
    \end{algorithmic}
\end{algorithm}

Another variational approach on quantum devices is QAOA (or Quantum Approximate Optimization Algorithm)~\cite{Zhou2018}, which can be seen as a generalization of VQE. First, one would define rotation parameters $\theta = [\theta_1, \ldots, \theta_d]$ and $\beta = [\beta_1, \ldots, \beta_d]$ together with the Ising Hamiltonian $H \in \mathbb{C}^{n\times n}$, and a mixing Hamiltonian $H_{X} \in \mathbb{C}^{n\times n} $ defined as a summations of $X$ Pauli operators. Then, one would construct the quantum state as
\begin{multline}
\ket{\psi(\theta, \beta)} = \exp(-i \beta_d H_X)\exp(-i\theta_d H) \cdots\\ \exp(-i \beta_1 H_X)\exp(-i \theta_1 H) \ket{0} = U(\theta, \beta) \ket{0}.    
\end{multline}
The variational approach would then aim at solving $\min_{\theta, \beta} \langle\psi(\theta, \beta)| H |\psi(\theta, \beta)\rangle$. Such optimization can be performed in a setting that uses a classical computer running an iterative algorithm to select $\theta, \beta$, and a quantum computer to compute information about $\langle\psi(\theta, \beta)| H |\psi(\theta, \beta)\rangle$ for given $\theta, \beta$ (e.g., its gradient). The algorithm outline of QAOA is reported in Algorithm \ref{algo:QAOA}.
	
\begin{algorithm}
		\caption{Outline of QAOA}\label{algo:QAOA} 
		\sf
		\begin{algorithmic}[1] 
			\small{\REQUIRE Hamiltonian $H$, mixer Hamiltonian $H_X$. Set $\theta = \theta_0$, $\beta = \beta_0$
			\WHILE{Error tolerance is unmet:}
			\STATE{Quantum part:}
			\begin{itemize}
            \item Form variational state $\ket{\psi({\theta, \beta})} = U(\theta, \beta)\ket{0}$
            \item Compute information about \mbox{$\lambda = \langle\psi(\theta, \beta)| H |\psi(\theta, \beta)\rangle$}
            \end{itemize}
            \STATE{Classical part:}
            \begin{itemize}
            \item Update $\theta, \beta$ via a classical optimization algorithm (e.g., COBYLA, SPSA, etc.)
            \item Compute the error metric
            \end{itemize}
            \ENDWHILE
            \RETURN $\lambda, \theta, \beta$}
\end{algorithmic}
\end{algorithm}

\subsection{Combinatorial Application 1: Active Investment Management, Portfolio Optimization}

To illustrate the VQE and the QAOA in the context of portfolio optimization we solve a combinatorial optimization problem in which we seek to allocate capital to a subset of $B=3$ assets selected from a larger investment universe with size $n=6$ \cite{barkoutsos2019improving}. In particular, we will solve the combinatorial problem:
\begin{equation}\label{eq:po}
\min_{x \in \{0, 1\}^n}  q x^{\intercal} \Sigma x - \mu^{\intercal} x, \quad \textrm{subject to: } 1^{\intercal} x = B,
\end{equation}
where we use the following notation:
\begin{itemize}
    \item $x \in \{0, 1\}^n$ denotes the vector of binary decision variables, which indicate which assets to pick ($x_i = 1$) and which not to pick ($x_i = 0$);
    \item $\mu \in \mathbb{R}^n$ defines the expected returns for the assets;
    \item $\Sigma \in \mathbb{R}^{n \times n}$ specifies the covariances between the assets;
    \item $q > 0$ controls the risk appetite of the decision maker;
    \item and $B$ denotes the budget, i.e. the number of assets to be selected out of $n$.
\end{itemize}
We also assume the following simplifications: \emph{i)} all assets have the same price (normalized to 1), \emph{ii)} the full budget $B$ has to be spent, i.e. one has to select exactly $B$ assets. 
\rev{With these assumptions the portfolio optimization problem corresponds to building a portfolio by selecting a subset of $B$ assets from the available $n$ assets and equally allocating capital to the $B$ assets.}
To map the problem~\eqref{eq:po} to a QUBO, the equality constraint $1^{\intercal} x = B$ is mapped to a penalty term $(1^{\intercal} x - B)^2$ which is scaled by a parameter and subtracted from the objective function. The resulting problem can be mapped to a Hamiltonian whose ground state corresponds to  the optimal solution. 

We use the statevector simulator in Qiskit Aer \cite{Qiskit} as backend and compare the results to a diagonalization of the Ising Hamiltonian which encodes the portfolio optimization problem.
The VQE and \rev{the} QAOA both produce variational states, which, when sampled from, result with high probability in an asset selection that respects the budget constraint \rev{as shown by the fact that the three most probable states all select three assets, see Tab.~\ref{tab:portfolio_optim}.
Furthermore, the selected states are either optimal or near optimal as seen by comparing them with the state resulting from the diagonalization in Tab.~\ref{tab:portfolio_optim}.
The probability to sample near optimal states with QAOA are lower than for VQE, see Tab.~\ref{tab:portfolio_optim}, which may indicate that deeper QAOA variational forms are needed \cite{Akshay2020} or that the COBYLA optimizer we used was trapped in a local minimum \cite{Shaydulin2019, Shaydulin2019b}.}
For such a small problem size the diagonalization runs in less time than simulations of the VQE and the QAOA.
Performing VQE and QAOA on quantum hardware would require even more time.
However, \rev{such classical brute force approach scales exponentially in the number of assets and, even for a few tens of assets ($n \sim 30-40$), we expect it not to be a practically viable approach.}

\begin{table}[t]
\caption{Comparison of the VQE solution, obtained with an $R_y$ variational form of depth 3, a depth $p=4$ QAOA solution, and a diagonalization of the Hamiltonian of the portfolio optimization problem. 
\rev{Assets selected shows the candidate solution where a $1$ in position $i$ indicates that asset $i$ was selected.}
The energy is the energy of the selected state and the probability shows the likelihood of sampling the selected assets from the quantum state created by the quantum algorithm.}
    \centering
    \sf
    \begin{tabular}{l r r} \toprule
         Assets selected & Energy & Probability \\ \toprule
         \multicolumn{3}{l}{Diagonalization} \\ \
         $[1\, 0\, 1\, 0\, 1\, 0]$ & -0.0036 & 1.0000 \\ \midrule
         \multicolumn{3}{l}{VQE} \\ \midrule
         $[0\, 1\, 0\, 0\, 1\, 1]$ & -0.0021 & 0.4431 \\
         $[0\, 1\, 0\, 1\, 1\, 0]$ & -0.0021 & 0.3660 \\
         $[1\, 1\, 0\, 0\, 1\, 0]$ & -0.0036 & 0.1154 \\\midrule
         \multicolumn{3}{l}{QAOA} \\ \midrule
         $[1\, 1\, 0\, 0\, 1\, 0]$ & -0.0036 & 0.0487 \\
         $[1\, 0\, 1\, 0\, 1\, 0]$ & -0.0036 & 0.0486 \\
         $[1\, 0\, 0\, 1\, 1\, 0]$ & -0.0031 & 0.0485 \\ \bottomrule
    \end{tabular}
    \label{tab:portfolio_optim}
\end{table}

\begin{figure}[t]
    \centering
    \includegraphics[width=\columnwidth]{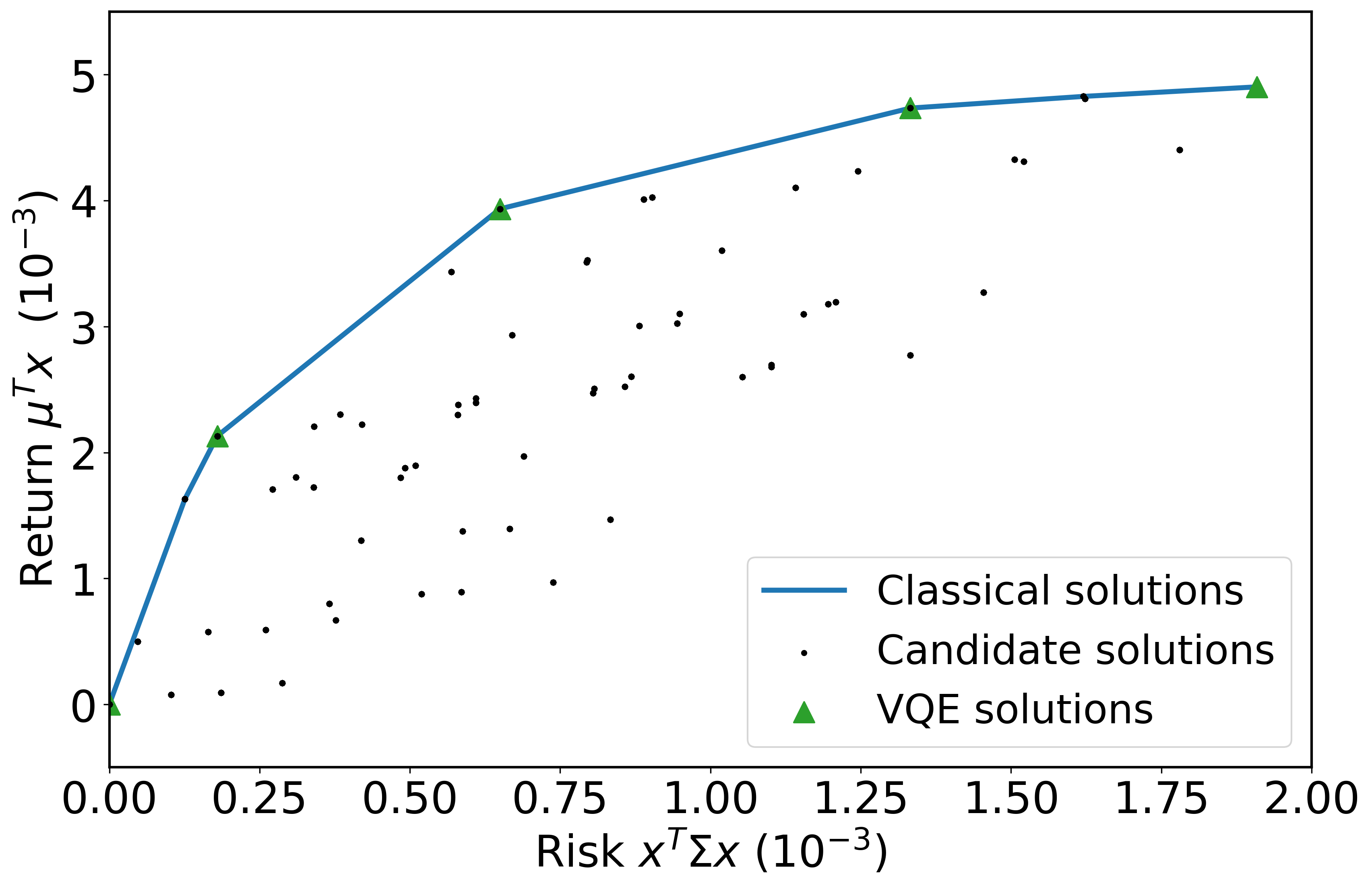}
    \caption{Reconstruction of the efficient frontier using \rev{an exhaustive search with exponential scaling} (blue line) and the VQE (green triangles). Both methods find the optimal portfolios, \rev{i.e. which maximize return and minimize risk}, out of the 64 possible asset combinations (black dots).
    \rev{Both quantum and classical algorithms are run several times with different risk-return trade-off values $q$ to find the efficient frontier.}}
    \label{fig:vqe_frontier}
\end{figure}

In a second example, we optimize the portfolio for different values of the risk-return trade-off parameter $q$ without the budget constraint.
We compare solutions obtained with VQE and solutions obtained from a classical \rev{exhaustive} search.
The most probable asset selections obtained from the state of the VQE closely follow the efficient frontier, therefore maximizing return and minimizing risk\rev{, see Fig.~\ref{fig:vqe_frontier}}.

\subsection{Combinatorial Application 2: Passive Investment Management, Portfolio Diversification}

In passive investment management one of the main challenges is to build a diverse portfolio with a relatively small number of assets that track the dynamics of a portfolio with a much larger number of assets. This
\textit{portfolio diversification} makes it possible to mimic the performance of an index (or a similarly large set of assets) with a limited budget, at limited transaction costs. The purchase of all assets in the index may be impractical for a number of reasons: the total of even a single round lot per asset may amount to more than the assets under management, the large scale of the index-tracking problem with integrality constraints may render the optimization problem difficult, and the transaction costs of the frequent rebalancing to adjust the positions to the weights in the index may render the approach expensive. Thus, a popular approach is to select a portfolio of $q$ assets that represent the market with $n$ assets, where $q$ is significantly smaller than $n$, but where the portfolio replicates the behavior of the underlying market. To determine how to group assets into $q$ clusters and how to determine which $q$ assets should represent the $q$ clusters amounts to solving a large-scale optimization problem.

As discussed in~\cite{Cornuejols2006}, we describe a mathematical model that clusters assets into groups of similar ones and selects one representative asset from each group to be included in the index fund portfolio. The model is based on the following data, which we will discuss in more detail later:
$$ 
\rho_{ij} = \textrm{similarity}\, \textrm{between}\, \textrm{stock}\, i \, \textrm{and}\, \textrm{stock}\, j. 
$$
For example, $\rho_{ii} = 1$, $\rho_{ij} \leq 1$ for $i \neq j$ and $\rho_{ij}$ is larger for more similar stocks. An example of this is the correlation between the returns of stocks $i$ and $j$. 
It allows for similarity measures between time-series beyond the covariance matrix. Consider, for instance, a company listed both in London and New York. Although both listings should be very similar, only parts of the time series of the prices of the two listings will overlap, because of the partial overlap of the times the markets open. Instead of covariance, one can consider, for example, dynamic time warping of~\cite{Berndt1994} as a measure of similarity between two time series, which allows for the fact that for some time periods, the data are captured by only one of the time series, while for others, both time series exhibit the similarity due to the parallel evolution of the stock price.

The problem that we are interested in solving is:
$$ (M) \quad f = \max_{x_{ij}, y_{j}} \,\, \sum_{i=1}^n \sum_{j=1}^n \rho_{ij} x_{ij} $$
subject to the clustering constraint:
$$ \sum_{j=1}^n y_j = q, $$
to consistency constraints:
\begin{eqnarray*}
&\displaystyle\sum_{j=1}^n x_{ij} = 1, & \forall\, i = 1,\ldots, n, \\
&x_{ij} \leq y_j, & \forall\, i = 1,\ldots, n; \, j = 1,\ldots, n, \\
&x_{jj} = y_j,& \forall\, j = 1,\ldots, n,
\end{eqnarray*}
and integral constraints:
$$ \quad x_{ij}, y_j \in\{0,1\}, \,\forall\, i = 1,\ldots, n; \, j = 1,\ldots, n. $$

The variables $y_j$ describe which stocks $j$ are in the index fund ($y_j = 1$ if $j$ is selected in the fund, $0$ otherwise). For each stock $i = 1,\dots,n$, the variable $x_{ij}$ indicates which stock $j$ in the index fund is most similar to $i$ ($x_{ij} = 1$ if $j$ is the most similar stock in the index fund, $0$ otherwise).

The first constraint selects $q$ stocks in the fund. The second constraint imposes that each stock $i$ has exactly one representative stock $j$ in the fund. The third and fourth constraints guarantee that stock $i$ can be represented by stock $j$ only if $j$ is in the fund. The objective of the model maximizes the similarity between the $n$ stocks and their representatives in the fund. Different cost functions can also be considered.

From $(M)$ one can construct a binary polynomial optimization with equality constraints only, by substituting the $x_{ij} \leq y_j$ inequality constraints with the equivalent equality constraints $x_{ij} (1- y_j) = 0$. Then the problem becomes:
\begin{subequations}
\begin{eqnarray}
\max_{x_{ij}, y_{j}}& \,\, \displaystyle \sum_{i=1}^n \displaystyle\sum_{j=1}^n \rho_{ij} x_{ij} &\\
\!\!\!\!\textrm{s. t.:} & \displaystyle\sum_{j=1}^n x_{ij} = 1, & \forall\,  i = 1,\ldots, n \\
& x_{ij} (1- y_j) = 0, & \forall\,  i = 1,\ldots, n, \nonumber\\
& &
\forall\, j = 1,\ldots, n, \\
& x_{jj} = y_j, & \forall\,  j = 1,\ldots, n.
\end{eqnarray}
\end{subequations}

We can now construct the Ising Hamiltonian (QUBO) by penalty methods (introducting a penalty coefficient $A$ for each equality constraint) as

\begin{multline}
\!\!\!H = \sum_{i=1}^n \sum_{j=1}^n \rho_{ij} x_{ij} + A\Big( \sum_{j=1}^n y_j - q\Big)^2 + \sum_{i=1}^n A\Big( \sum_{j=1}^n x_{ij} - 1\Big)^2 \\ + \sum_{j=1}^n A (x_{jj}-y_j)^2 +\sum_{i=1}^n \sum_{j=1}^n A \left(x_{ij} (1- y_j)\right).    
\end{multline}

For the simulation in Qiskit, we use three assets ($n = 3$) and two clusers ($q = 2$), this leads to a $12$-qubit Hamiltonian. We solve the problem classical with CPLEX and on the quantum computer with VQE (with depth 7 and full entanglement). 

\begin{figure}[t]
    \centering
    \includegraphics[width=0.9\columnwidth]{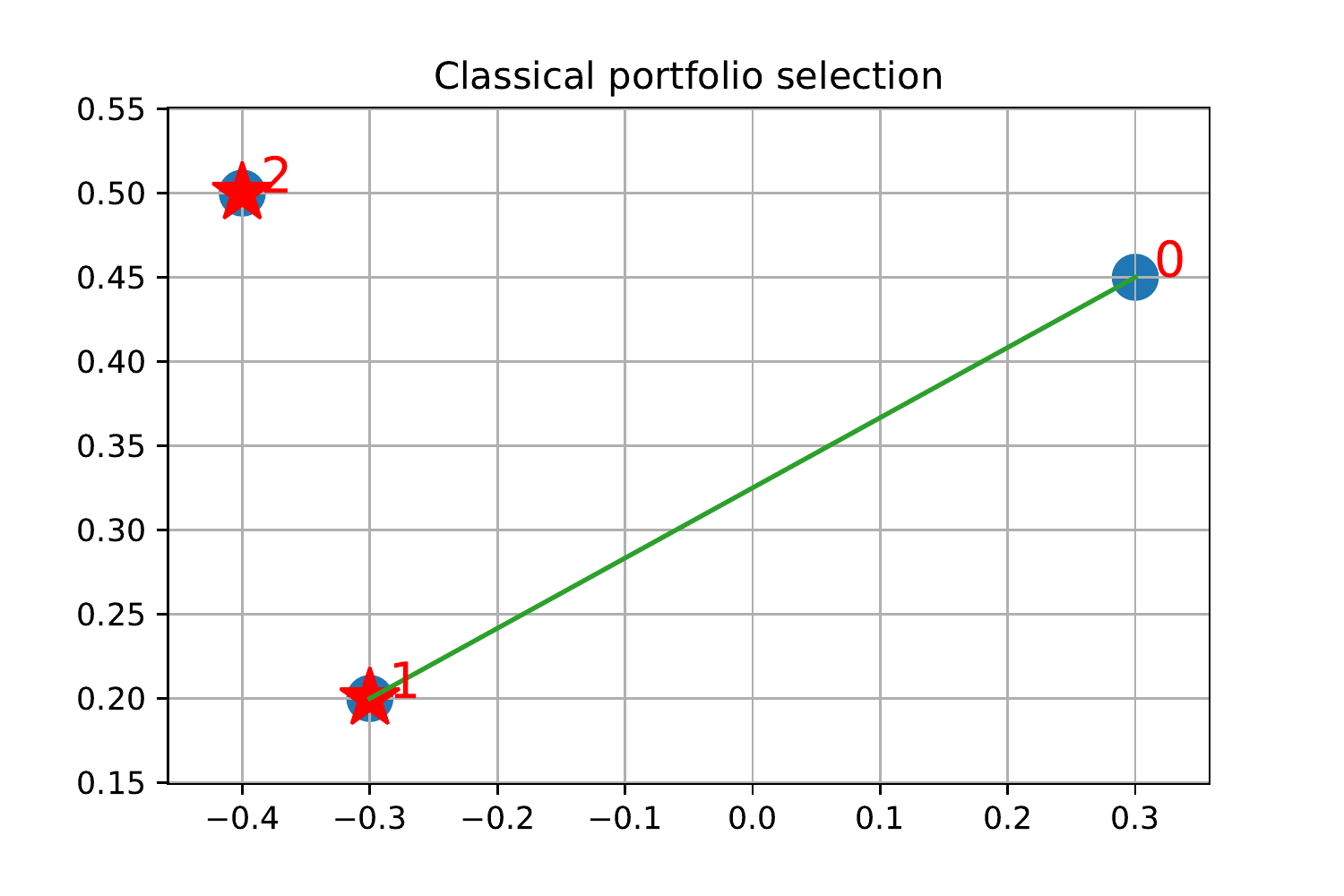}
    \includegraphics[width=0.9\columnwidth]{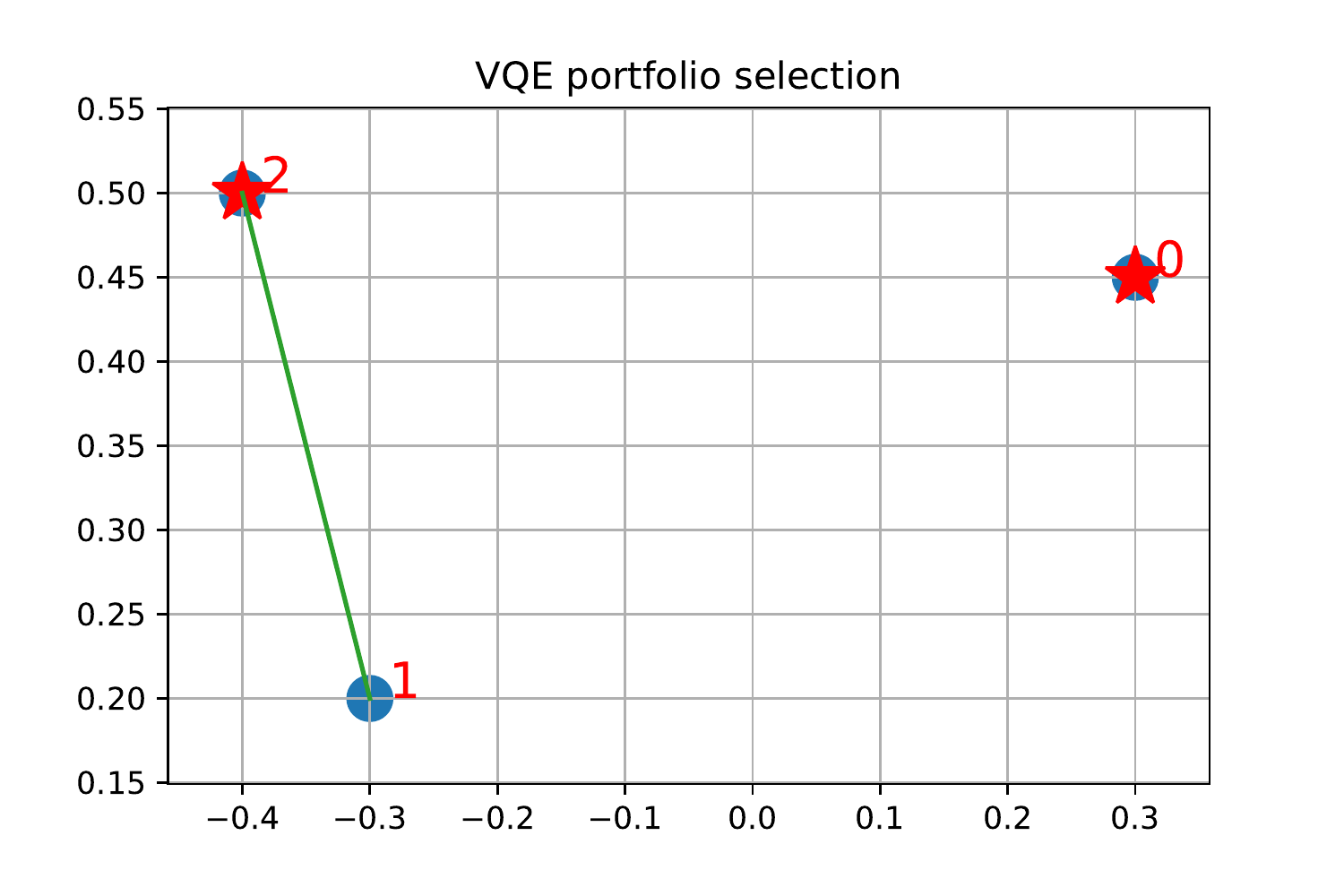}
    \caption{Portfolio selection for classical and quantum algorithm. \rev{The stocks are represented in an arbitrary two dimensional plane for viualization purposes. The selected stocks are indicated with red stars, while the green link indicates the association of the unselected stock to a selected one. Classical and quantum solution are different but fare very similarly in terms of benefit. }}
    \label{fig:pd}
\end{figure}

In Figure~\ref{fig:pd}, we report the results that we obtain \rev{conveniently (and arbitrarily) displayed in a two dimensional graph for visualization purposes}. Solution shows the selected stocks via the stars and in green the links (via similarities) with other stocks that are represented in the fund by the linked stock.  As we see, both \rev{for} classical and quantum, we can find a feasible solution for our diversification (\rev{clustering and selecting two stocks, while associating the third to a selected one}), although the classical algorithm here finds a slightly better solution (\rev{the classical benefit fares at $2.001$, while the quantum one at $2.000$}). This is reasonable for such small problem instance, since the classical solver  \rev{can be run to find the exact solution}, while VQE is a heuristic \rev{and may find less optimal solutions}, \rev{but this might not be the case when the classical solver will not be able to run at optimality for larger size problems}.

		{\begin{algorithm}[t]
		\caption{\threeadmm~mixed-binary heuristic}\label{algo:threeblock}
		\sf
		\begin{algorithmic}[1]
			\small{\REQUIRE Initial choice of $x_0, \bar{x}_0, y_0, \lambda_0$. Choice of $\varrho, \beta, c >0$, tolerance $\epsilon>0$, and maximum number of iterations $K_{\text{max}}$.
				\WHILE {$k<K_{\text{max}}$ \AND $\|A_0 x_k-
					A_1 \bar{x}_k - y_k\| < \epsilon,$}
				\STATE First block update (QUBO) on the \textbf{quantum device}:
				\begin{align}
				x_k = & \argmin_{x \in \{0,1\}^n} \,\, q (x) + \frac{c}{2}\|Gx - b\|^2_2 + \nonumber\\
				& + \lambda_{k-1}^{\transp} A_0 x + \frac{\varrho}{2}\|A_0 x + A_1 \bar{x}_{k-1} - y_{k-1} \|^2 \label{eq:QUBO}
			\end{align}
				\STATE Second block update (Convex) on the \textbf{classical device}:
				\begin{align}
				\bar{x}_k & = \argmin_{\bar{x} \in \mathbb{R}^m } \,\, f_1(\bar{x}) + \lambda_{k-1}^{\transp} A_1 \bar{x} + \nonumber\\
				&  \frac{\varrho}{2}\|A_0 x_{k} + A_1 \bar{x} - {y}_{k-1}\|^2
				\end{align}
				\STATE Third block update (Convex+quadratic) on the \textbf{classical device}:
				$$
				y_k = \argmin_{y \in \mathbb{R}^n } \,\, \frac{\beta}{2}\|y\|^2_2 - \lambda_{k-1}^{\transp} y + \frac{\varrho}{2}\|A_0 x_{k} + A_1 \bar{x}_{k} - {y}\|^2
				$$
				\STATE Dual variable update on the \textbf{classical device}:
				$$
				\lambda_k = \lambda_{k-1} + \varrho(A_0 x_{k} + A_1 \bar{x}_k - y_k)
				$$
								\STATE Compute merit value:
\begin{align}
				\eta_k & = q(x_k)+\phi(\bar{x}_k) + \nonumber\\
				& + \mu (\max (g(x_k), 0) + \max(l(x_k, \bar{x}_k), 0))
\end{align}
				\ENDWHILE
			 \RETURN $x_{{k}^*}, \bar{x}_{{k}^*}, y_{{k}^*},$ with ${{k}^*} = \min_k \eta_k.$}
		
	\end{algorithmic}\end{algorithm}}

\begin{figure*}[t]
\centering
\includegraphics[width=0.85\textwidth]{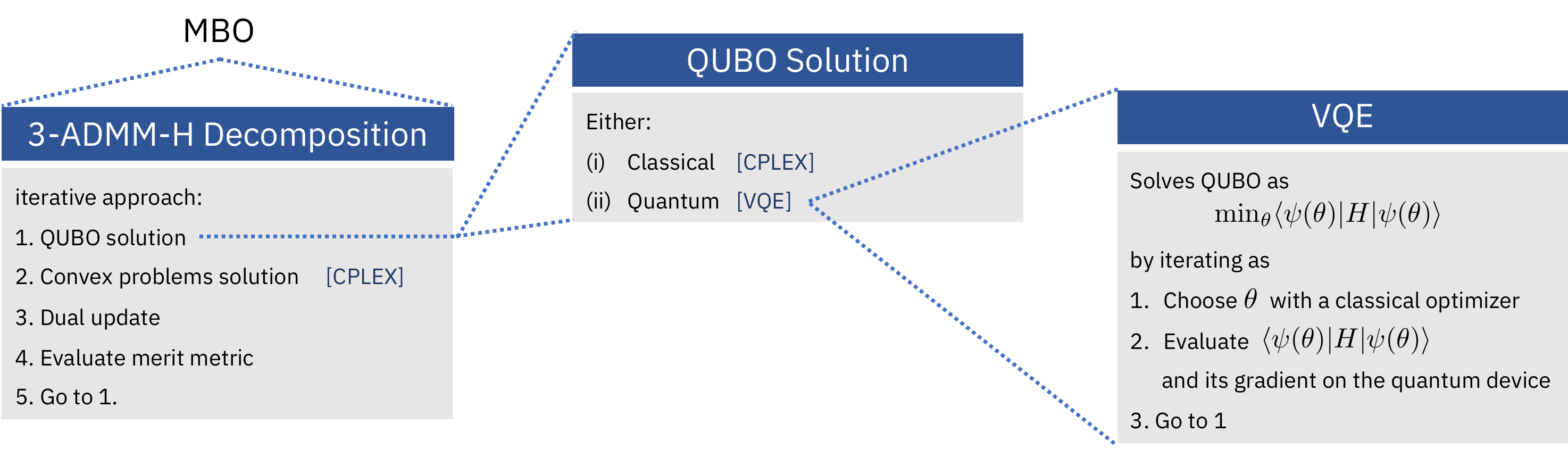}
\caption{Illustration diagram of the \threeadmm approach. 
There are two nested loops for the selected implementation, specifically the outer ADMM loop, and the inner VQE loop (Note VQE can be substituted via QAOA seamlessly).}
\label{fig.appr}
\end{figure*}

\subsection{Multi-block ADMM heuristic for Mixed-Binary Optimization}\label{algo:ADMM}

We move on to Mixed-Binary Optimization (MBO) formulations. In a general MBO problem, the decision maker faces binary and continuous decisions, subject to equality and inequality constraints. \rev{MBO formulations enable to tackle finance problems, such as the combinatorial auction problem, which is the scope of the Section \ref{sec:sim}.}

\rev{In order to introduce solvers for MBO, we} consider the following reference \rev{MBO} problem $(P)$:
\begin{subequations}\label{eq:MBO}
\begin{eqnarray}
\min_{x \in \mathcal{X},u\in\mathcal{U} \subseteq \mathbb{R}^l } \ &&  q(x) + \varphi(u)  \\
\mathrm{s.t.:~} && G x = b, \quad  g(x) \leq 0 \\
&&  \ell(x, u) \leq 0,
\end{eqnarray}
\end{subequations}
with the corresponding functional assumptions.

\begin{assumption}\label{as:1} The following assumptions hold:
\begin{itemize}
	\item Function $q: \mathbb{R}^n \to \mathbb{R}$ is quadratic, i.e., $q(x) = x^{\transp} Q x + a^{\transp} x$ for a given symmetric squared matrix $Q \in \mathbb{R}^n \times \mathbb{R}^n, Q = Q^{\transp}$, and vector $a \in \mathbb{R}^n$;
	\item The set $\mathcal{X} = \{0,1\}^n = \{x_{(i)} (1-x_{(i)}) = 0, \forall i\}$ enforces the binary constraints;
	\item Matrix $G\in\mathbb{R}^n \times \mathbb{R}^{n'}$, vector $b \in \mathbb{R}^{n'}$, and function $g: \mathbb{R}^n \to \mathbb{R}$ is convex;
	\item Function $\varphi: \mathbb{R}^l \to \mathbb{R}$ is convex and $\mathcal{U}$ is a convex set;
	\item Function $\ell: \mathbb{R}^n\times  \mathbb{R}^l \to \mathbb{R}$ is \emph{jointly} convex in $x, u$.
\end{itemize}
\end{assumption}

In order to solve MBO problems, \cite{gambella2020multi} proposed heuristics for $(P)$ based on the Alternating Direction Method of Multipliers (ADMM) ~\cite{boyd2011distributed}. ADMM is an operator splitting algorithm with a long history in convex optimization, and it is known to have residual, objective and dual variable convergence properties, provided that convexity assumptions are holding \cite{boyd2011distributed}.

The method of \cite{gambella2020multi} (referred to as \threeadmm, and displayed in Figure \ref{fig.appr}) leverages the ADMM operator-splitting procedure to devise a decomposition for certain classes of MBOs into:
\begin{itemize}
	\item a QUBO subproblem to be solved by on the quantum device via variational algorithms, such as VQE or QAOA, described in Section \ref{sec:alg-qubo};
	\item a continuous convex constrained subproblem, which can be efficiently solved with classical optimization solvers \cite{boyd2004convex}. 
\end{itemize} 
Algorithm~\ref{algo:threeblock} reports the \threeadmm~algorithm, along with stopping criteria and evaluation metrics. A comprehensive discussion on the conditions for convergence, feasibility and optimality of \threeadmm~is out of the scope of the present document and can be found in \cite{gambella2020multi}. Combinatorial auction (A) belongs to the class of MBOs represented by (P) and can be solved by \threeadmm. Simulations on representative instances are conducted in Section \ref{sec:sim}.

\subsection{Combinatorial Application 3: Auctions}\label{sec:sim}

Both governments and private issuers finance its activities, in part, by the sale of marketable securities. The issuer often uses an auction process to sell such marketable securities and determine their parameters (such as yield). For example, the United States (U.S.) treasury issued over \$10T (ten trillion U.S. dollars) in securities in 2018.
Many further auction mechanisms abound in electrical energy markets, pollution management, and within airport operations (airport landing slots). 
Some of these auctions may be combinatorial, in the sense  that the value that a bidder has for a set of items may not be the sum of the values that he has for individual items. It may be more or it may be less. 

Combinatorial auctions allow the bidders to submit bids on subsets (combinations) of items.
Specifically, let $M = \{1, 2, \ldots , m\}$ be the set of items that the auctioneer has to sell. A bid is a pair $B_j =(S_j,p_j)$ where $S_j \subseteq M$ is a non empty set of items and $p_j$ is the price offer for this set. Suppose that the auctioneer has received $n$ bids $B_1, B_2, \ldots, B_n$. How should the auctioneer determine the winners in order to maximize his revenue? This can be formulated as an integer program. To render the problem a bit more interesting, we consider the case in which (as in some auctions) there are multiple indistinguishable units of each item for sale. A bid in this setting is defined as $B_j = (\lambda_{1}^j, \lambda_{2}^j,\ldots,\lambda_{m}^j;p_j)$ where $\lambda_{i}^j$ is the desired number of units of item $i$ and $p_j$ is the price offer. Let $x_j$ be a binary variable that takes the value $1$ if bid $B_j$ wins, and $0$ if it loses. The auctioneer maximizes his revenue by solving the integer program (A):
\begin{eqnarray}
\max_{x_j} && \quad \sum_{j=1}^n p_{j} x_{j} \\
\text{s.t.:} && \sum_{j: i \in S_j} \lambda_{i}^j x_j \leq u_i, \quad \textrm{for } i = 1, \ldots, m \label{cons:auc-cap}\\
&& x_j \in \{0,1\}, \quad \textrm{for } j = 1, \ldots, n,
\end{eqnarray}
where $u_i$ is the number of units of item $i$ for sale~\cite{Cornuejols2006}.

The presence of inequality constraints \eqref{cons:auc-cap} makes a reformulation of (A) into a QUBO not possible, hence the VQE algorithm described in Section \ref{sec:alg-qubo} is not directly applicable. We here report results obtained by solving (A) with the \threeadmm~heuristic described in Section \ref{algo:ADMM}.

For simulation purposes, an instance with $m = 3$ items and $n = 16$ bids with randomly generated profits has been created. For each item, $6$ units are available. The number of units of the items in each bid has been randomly sampled from the interval $[1, 6]$. This means that not all bids are necessarily feasible if more than one object is in the bid. Because the decision of accepting a bid $j$ is a binary decision $x_j$, the number of bids is the number of qubits the algorithm \threeadmm~necessitates.  
The optimal solution, found by solving Problem (A) via the classical optimization solver IBM ILOG CPLEX, consist of accepting bids $B_0 = \{0\}, B_1=\{1\}, B_4=\{1, 2\}$ with a profit of $24$. 
The \threeadmm~algorithm has been tested on the instance by choosing VQE as quantum solver in Qiskit Aqua and Constrained Optimization By Linear Approximation (COBYLA) \cite{gomez2013advances} as classical VQE optimizer with $20$ maximum iterations. The \texttt{qasm\_simulator} has been used as Qiskit Aer backend for the simulations on quantum devices. 
The ADMM parameters $\rho$ and $\beta$ have been set to $12$ and $11$ respectively: this is to leverage the convergence properties described in \cite{gambella2020multi} for $\rho>\beta$. When run on classical devices, the first block update is performed with the CPLEX solver.

\begin{figure}[ht!]
	\centering
\begin{subfigure}[b]{\columnwidth}
\includegraphics[width=\linewidth]{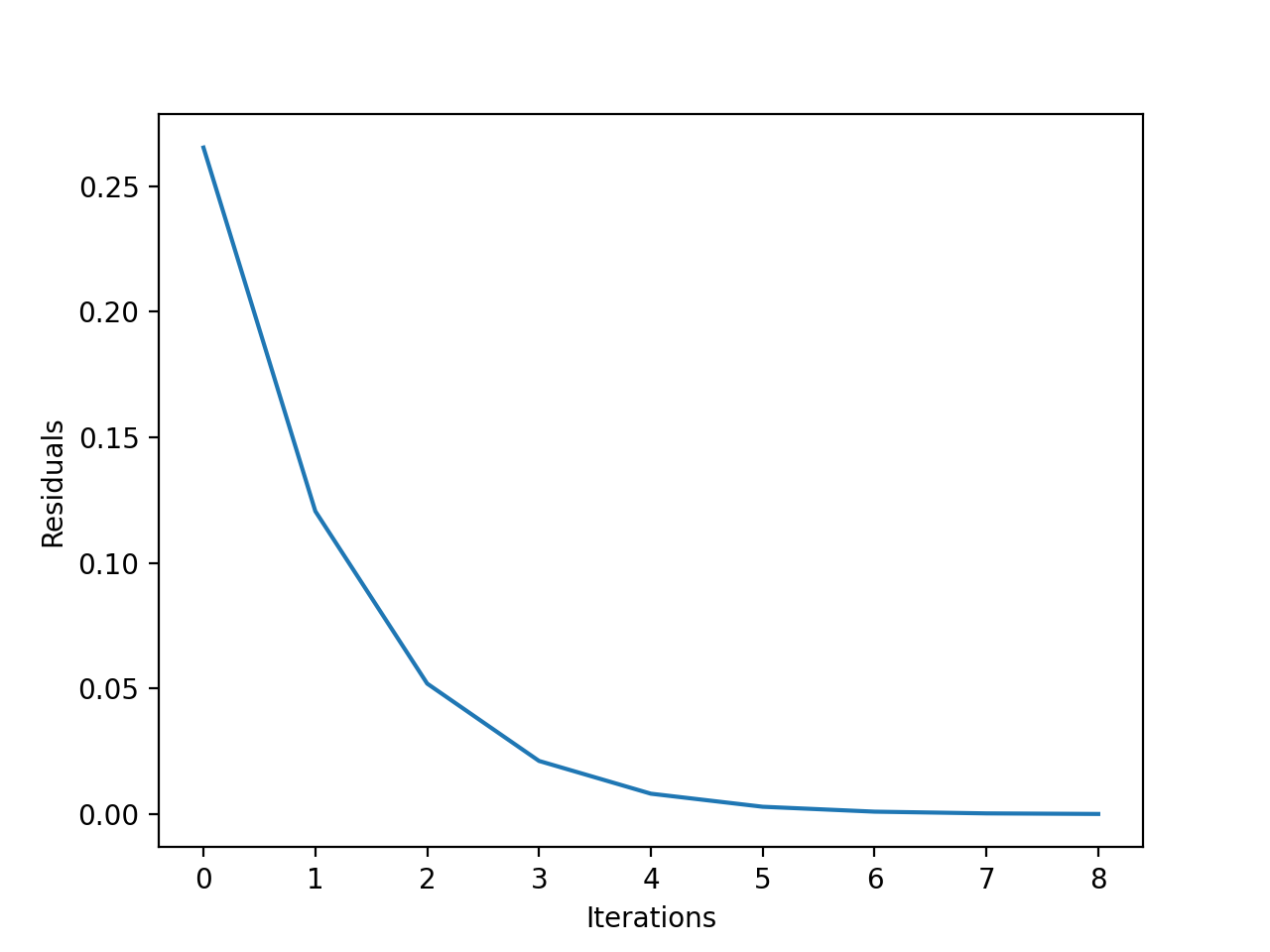}
		\caption{}
		\label{fig:admm-c}
		\end{subfigure}\\%
\begin{subfigure}[b]{\columnwidth}
		\includegraphics[width=\linewidth]{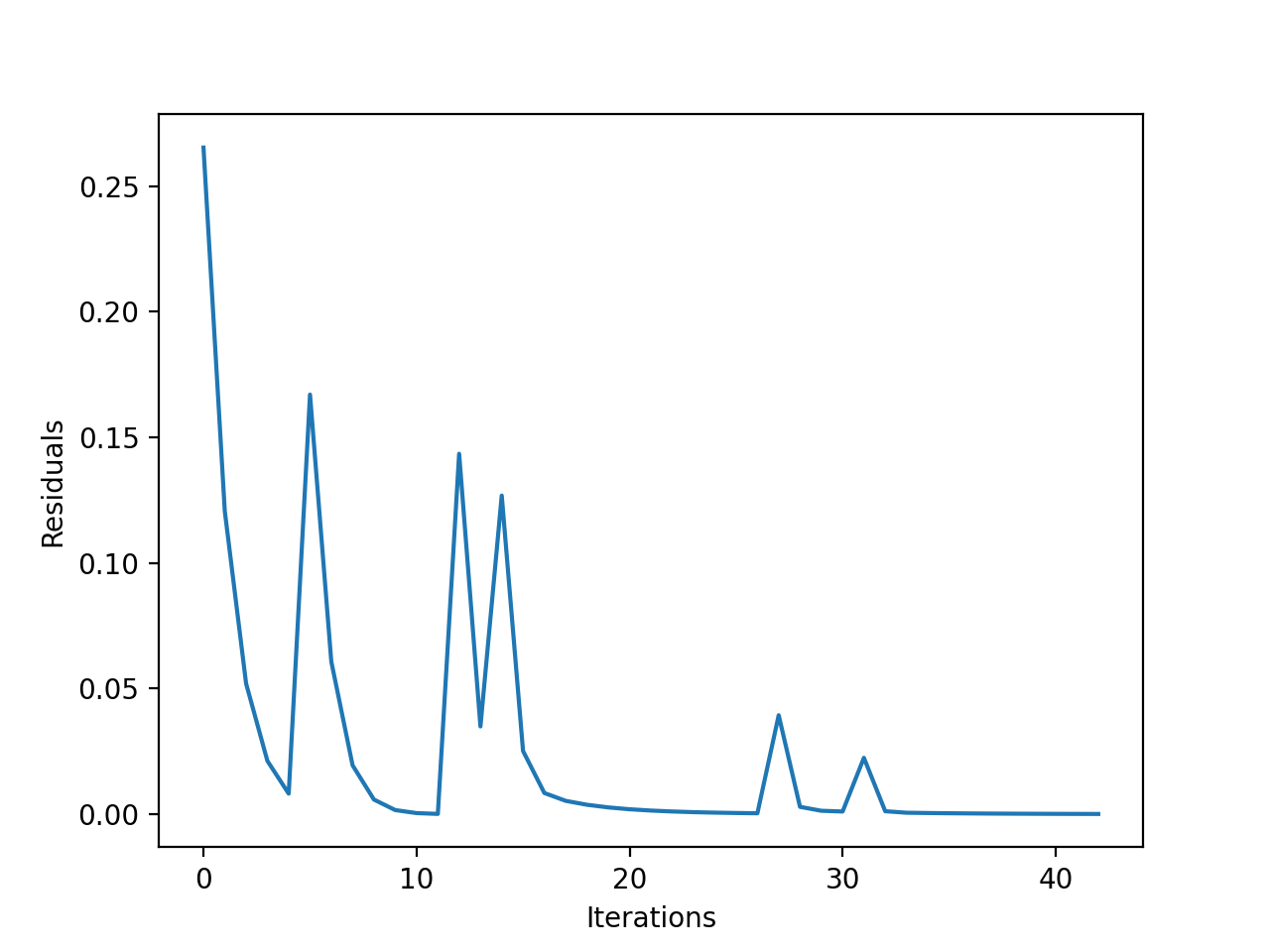}
				\caption{}
		\label{fig:admm-q}
		\end{subfigure}%
	\caption{\rev{Convergence of the residuals for \threeadmm~algorithm, in the classical (a) and quantum (b) simulations. The simulation with quantum QUBO solver shows that, even with an inexact QUBO solver, \threeadmm~can reach convergence in a finite, and relatively low, number of iterations. }}
	\label{fig:admm}
\end{figure}

The \threeadmm~solution with CPLEX as QUBO solver  is  $B_1 = \{1\}, B_3=\{0, 2\}, B_4=\{1, 2\},$ with a profit of $27$, and a violation of constraints \eqref{cons:auc-cap} of $2$. Setting VQE as QUBO solver makes \threeadmm~converge to the same solution in $43$ iterations. The residuals \mbox{$r_k = A_0 x_k-A_1 \bar{x}_k - y_k$} are reported for the classical and quantum simulations, in Figures \ref{fig:admm-c} and \ref{fig:admm-q}. \rev{Residuals are not guaranteed to decrease in each \threeadmm~iteration. The convergence guarantees of \threeadmm~are not valid for inexact QUBO solvers, such as the currently available quantum algorithms. However, the convergence curves show that \threeadmm~terminates in a finite number of iterations, even when the QUBO solver is inexact. Hence, \threeadmm~exhibits a certain degree of tolerance to inexact computations. This corroborates the empirical findings of \cite{gambella2020multi} on packing problems.}

The \threeadmm~algorithm proposes a decomposition of a mixed-binary optimization problem, in which the most computationally demanding part is solving the QUBO subproblem \eqref{eq:QUBO}.
The advantage of using \threeadmm~algorithm over classical optimization solvers, such as CPLEX, lies in leveraging quantum algorithm to tackle QUBO subproblems.

\section{Machine Learning}
\label{sec:ml}

Finally, in this section we discuss machine learning problems where quantum algorithms may demonstrate an advantage. 

Machine learning focuses on finding relations in data and building assumptions around them for (i) prediction by anticipating future events from historic data, or to (ii) classify data by dividing an end result into different categories, or to (iii) find patterns by the discovery of regularities or anomalies in data. 
In finance, such approaches are important in many financial problems that deal with uncertainty in the future evolution of asset prices and risk. For example, the investment management strategies and optimization in the previous section make use of the estimation of future risks and asset prices that can be obtained from the output of machine learning algorithms. Banks can estimate the risk level of their customers' loans by credit scoring which can be formulated as classification \cite{Lessmann2015} and/or regression problem based on the rich features of customers, such as, age, salary, historical payment, micro and macro economic indicators and so on. Financial institutions can also detect frauds by finding patterns that deviates greatly from normal behavior by classification and/or anomaly detection \cite{Li_2012,ram2018}. Such machine learning tasks are known to face \textit{the curse of dimensionality} as there are much more features available to model customers. \rev{Principal} component analysis (PCA) and variational autoencoder (VAE) \cite{doersch2016tutorial} are some of the popular methods for dimensionality reduction when dealing with high-dimensional features. 

To summarize machine learning problems at each stage of the customer life cycle (see Figure \ref{fig:benefits-pipeline}):

\begin{enumerate}
\item
\textbf{Customer identification (and scoring)}
Refine customer rating and segmentation to improve KYC (Know Your Customer) and avoid non-compliance annual penalties, that have grown to \$ 8.3 Million annually by 2018 \cite{Realwire2019}.
\item
\textbf{Financial products}
Increase credit scoring realism to improve customer targeting and align product offering: 
\begin{itemize}
\item
Avoid NPL (Non-Performing Loan) costs that are still increasing in many European economies \cite{Datatopics2019}.
\item
Increase customer interaction with the bank resulting in more profits. 
\item
Improve recommendation techniques to target customers prone to accept them.
\end{itemize}
\item
\textbf{Monitor transactions}
Improve suspicious transaction signals to decrease false alerts, that today cost \$ 4 trillion due to 75\% to 90\% false positives in AML (Anti Money Laundering) and credit card fraud alerts.
\item
\textbf{Customer retention}
Maximize engagement to avoid customer churn rates to new entrants, 25\% of SMBs (Small-Medium Business) are turning to FinTech companies for ease and speed of completing loan applications \cite{Pymnts2017}. 
\end{enumerate}

\rev{Overall, Artificial Intelligence (AI) and Machine Learning (ML) are of deep interest for financial institutions, with a current global spending in the banking industry worth of \$ 3.3 Billion in 2018\cite{Emerj2019} with the hopes to build better classification models that will improve customer service in external facing and internal activities. IDC reported that of global spending on AI worth \$ 50.1 Billion in 2020, which is expected to double in four years, Banking is one of the largest industries spending the most in AI/ML solutions for fraud analysis, investigation, program advisors and recommendation systems~\cite{IDC2020}.}

In the following, we discuss how quantum-enhanced feature space can be used in a simple task of binary classification that can be applied to financial applications, such as, fraud detection (for transaction monitoring) and credit risk scoring (for customer identification). There are many other tasks addressable by quantum machine learning techniques, e.g., see~\cite{Orus2019,Biamonte2017} for more tasks and applicable quantum techniques. We focus in supervised learning using Support Vector Machine (SVM): we have access to labeled training data $\mathit{S}$ to classify test data $\mathit{T}$ and labels of unseen future data (with assumption that all data come from the same underlying distribution). 

Assume that we are given the training data $\mathit{S} = \left\{ (\ru{x}_1, y_1), (\ru{x}_2, y_2), \ldots, (\ru{x}_{m_S}, y_{m_S}) \right\}$, where each $\ru{x}_i \in \mathbb{R}^d$ and $y_i \in \{-1, 1\}$. The goal of learning a binary classifier from $\mathit{S}$ is to construct a function $f(\ru{x})$ so that $f(\ru{x})y_i > 0$. The simplest form of such function is a linear classifier $f(\ru{x}) = \ru{w}^{\intercal}\ru{x} + b$, where $(\ru{w}, b) \in \mathbb{R}^{d+1}$. $\mathit{S}$ is called linearly separable if there is a $(\ru{w}, b) \in \mathbb{R}^{d+1}$ satisfying $f(\ru{x})y_i > 0$. Such function, if exists, can be found by solving an optimization problem known as Hard-SVM. 

In general, the dataset may not be linearly separable. In such case we can still find a classifier that predicts the training dataset with some error margin. The fomulation is know as Soft-SVM, as shown below, and can be solved efficiently by techniques such as Stochastic Gradient Descent (SGD). 

\begin{eqnarray*}
(\ru{w}_0, b_0) &=& \argmin_{(\ru{w}, b)} \|\ru{w}\|^2 + C \sum_{i=1}^{m_S} \epsilon_i \\
\mbox{subject to} &:& y_j~(\ru{w}^{\intercal}\ru{x}_j + b) \ge 1 - \epsilon_j~\mbox{for}~j\in [m_S]\\
&& \epsilon_j \ge 0~\mbox{for}~j\in[m_S]
\end{eqnarray*}

The slack variables $\{\epsilon_i\}$ determine the quality of the classifier: the closer they are to zero the better the classifier. For this purpose, we can embed the data $\{\ru{x}_j\}$ into a larger space by pre-processing the data, namely, by finding a map $\ru{x}: \Phi(\ru{x}) \in \mathbb{R}^{n}$ for $n > d$. The classifier $f(\ru{x})$ is now defined as $ f(\ru{x}) = \ru{w}^{\intercal} {\Phi(\ru{x})} + b$. When $\Phi(\ru{x})$ is an embedding of data non-linearly to quantum state $\left|\Phi(\ru{x})\right>$ then we can use quantum-enhanced feature space for the classifier. There are two techniques to construct such a quantum-enhanced feature space that may lead to a Quantum Advantage: Variational Quantum Classification (VQC) and Quantum Kernel Estimation (QKE).

The VQC is similar to SGD for finding the best hyperplane $(\ru{w}, b)$ that linearly separates the embedded data. At VQC, the data $\ru{x} \in \mathbb{R}^d$ is mapped to (pure) quantum state by the feature map circuit $\ru{U}_{\Phi(\ru{x})}$ that realizes $\Phi(\ru{x})$. This means, that conditioned on the data $\ru{x}$, we apply the circuit $\ru{U}_{\Phi(\ru{x})}$ to the $n$-qubit all-zero state $\left|0_n\right>$ to obtain the quantum state $\left|\Phi(\ru{x})\right>$. A short-depth quantum circuit $\ru{W}(\ru{\theta})$ is then applied to the quantum state, where $\ru{\theta}$ is the hyperparameter set of the quantum circuit that can be learned from the training data. Finding the circuit $\ru{W}(\ru{\theta})$ is akin to finding the separating hyperplane $(\ru{w},b)$ in the Hard-SVM and Soft-SVM, with the path to Quantum Advantage stems from the fact that there is no efficient classical procedure to realize the feature map $\Phi(\ru{x})$. While the size of the hyperparameter set $\ru{\theta}$ is polynomial in the number of qubits and can be tuned with variational methods similar to Algorithm~\ref{algo:VQE}~and~\ref{algo:QAOA}, it controls an exponentially large space of the feature map. The binary decision is obtained by measuring the quantum state in the computational basis to obtain $\ru{z} \in \left\{0,1\right\}^n$, and linearly combining the measurement results, say with $\ru{g} = \sum_{\ru{z} \in \left\{0,1\right\}^n} g(\ru{z}) \left|\ru{z}\right>\left<\ru{z}\right|$, where $g(\cdot) \in \{-1, 1\}$. 

\begin{figure}[t!]
\centering
\includegraphics[width=0.9\columnwidth]{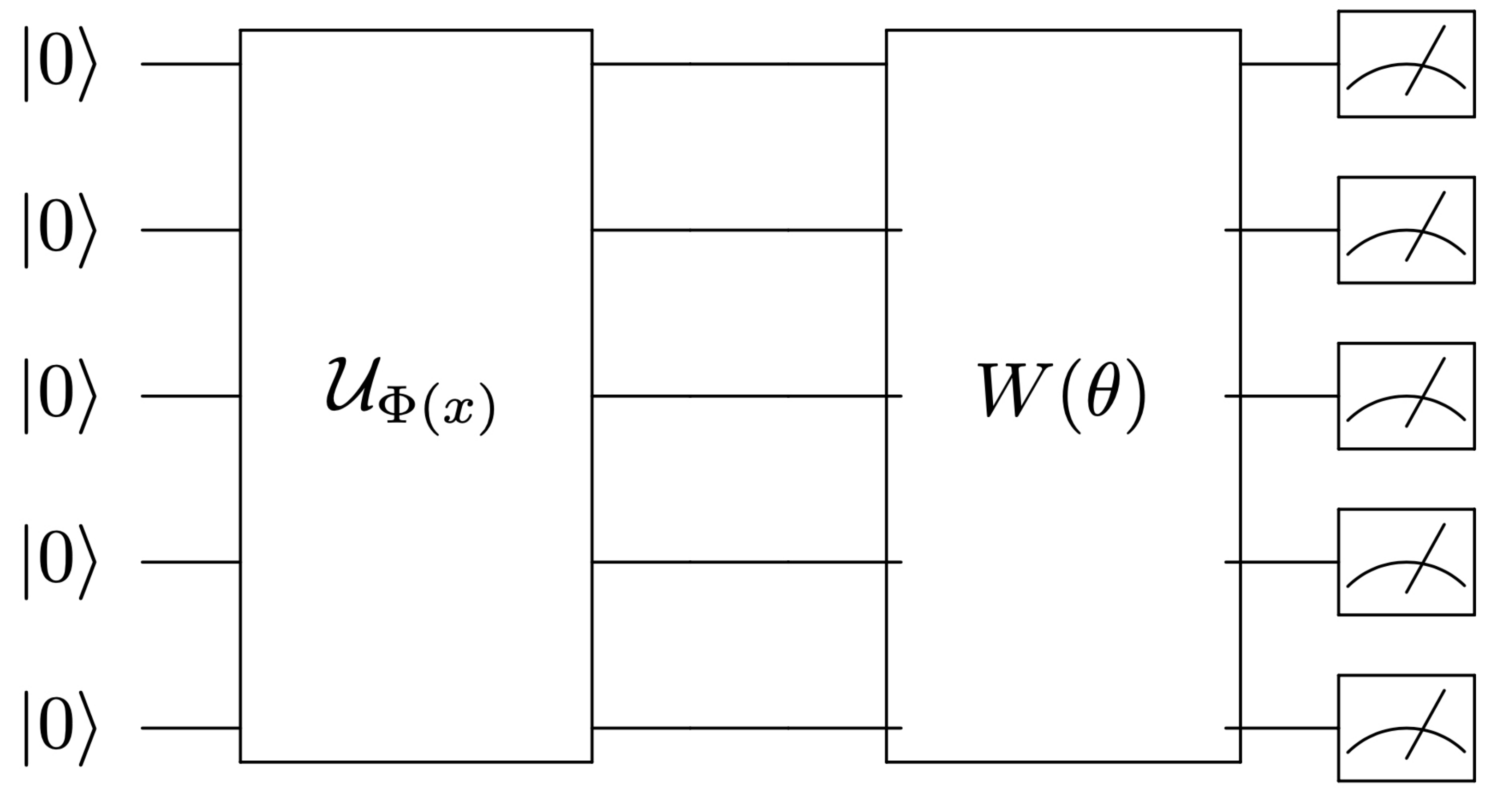}
	\caption{Quantum circuit for Variational Quantum Classifier (VQC) that consists of fixed quantum feature mapping $\mathcal{U}_{\Phi(\ru{x})}$ and the separator $\ru{W}(\ru{\theta})$ trained with variational methods.}
	\label{fig:vqc}
\end{figure}

A quantum circuit that realizes the quantum feature map as well as the variational classifier is shown in Fig.~\ref{fig:vqc}. We can see that the probability of observing $\ru{z}$ is given as 
\begin{equation*}
\left| \left<\ru{z}\right| \ru{W}(\ru{\theta})\left|\Phi(\ru{x})\right> \right|^2 = \left<\Phi(\ru{x})\right| \ru{W}^\dag(\ru{\theta})\left|\ru{z}\right>\left<\ru{z}\right|\ru{W}(\ru{\theta})\left|\Phi(\ru{x})\right>.
\end{equation*}
By linear combination of the measurement results $\ru{z}$ with $\ru{g}$, we can obtain the function $f(\ru{x})$ as below, which resembles the linear classifier $f(\ru{x}) = \ru{w}^{\intercal} \ru{\Phi}(\ru{x}) + b$:
\begin{equation*}
    f(\ru{x}) = \left<\Phi(\ru{x})\right| \ru{W}^\dag(\ru{\theta})~\ru{g}~\ru{W}(\ru{\theta})\left|\Phi(\ru{x})\right> + b.
\end{equation*}
The predicted label of $f(\ru{x})$ is then simply its sign. The hyperplane $(\ru{w}, b)$ is now parametrized by $\ru{\theta}$. The $i$-th element of $\ru{w}(\ru{\theta})$ is $w_i(\ru{\theta}) = \tr\left(\ru{W}^\dag(\ru{\theta}) \ru{g} \ru{W}(\ru{\theta})\ru{P}_i \right)$, where $\ru{P}_i$ is a diagonal matrix whose elements are all zeros except at the $i$-th row and column which is $1$, and the $i$-th element of $\ru{\Phi}(\ru{x})$ is $\Phi_i(\ru{x}) = \left<\ru{\Phi}(\ru{x})\right|\ru{P}_i\left|\ru{\Phi}(\ru{x})\right>$. 

Learning the best $\ru{\theta}$ can be obtained by minimizing the empirical risk $R(\ru{\theta})$ with regards to the training data $\mathit{S} = \left\{ (\ru{x}_1, y_1), (\ru{x}_2, y_2), \ldots, (\ru{x}_{m_S}, y_{m_S}) \right\}$. Namely, the empirical risk (or, cost function) to be minimized is 
\begin{equation}
    R(\ru{\theta}) = \frac{1}{|S|} \sum_{i \in [m_\mathit{S}]} \left|f(\ru{x}_i) - y_i \right|.\label{eq:emr}
\end{equation}
The above empirical risk can then be approximated with a continuous function using \textit{sigmoid} function as detailed in \cite{Havl_ek_2019}. This enables applying variational methods as in Algorithm~\ref{algo:VQE}~and~\ref{algo:QAOA} with stochastic gradient descent algorithms (such as, COBYLA or SPSA) for tuning $\ru{\theta}$ to minimize the cost function.  

The binary classification with VQC now follows from first training the classifier to learn the best $\ru{\theta}^*$, that minimizes the empirical risk $R(\ru{\theta})$, to obtain $(\ru{w}(\ru{\theta}^*), b^*)$. This can be done with Algorithm~\ref{algo:VQE} with the Hamiltonian replaced by the empirical risk. The classification against unseen data $\ru{x}$ is then performed according to the classifier function $f(\ru{x})$ with $(\ru{w}(\ru{\theta}^*), b^*)$. Both training and classification need to be repeated for multiple times (or, shots) due to the probabilistic nature of quantum computation. The former may need significant number of shots proportional to the size of $\mathit{S}$ but it can be performed in batch offline. One the other hand, the latter needs much less number of shots, and may be performed online (or, near real time) as long as the quantum feature map for non-linear embedding can be computed efficiently. 

In the conventional SVM, there are many known methods of non-linear embedding of data $\ru{x}: \ru{\Phi}(\ru{x}) \in \mathbb{R}^{n}$ for $n > d$, such as, Polynomial-SVMs. For example, in a Polynomial-SVM, the 2-dimensional data $(x_1, x_2)$ can be embedded into a 3-dimensional $(z_1, z_2, z_3)$ such that, $z_1 = x_1^2$, $z_2 = \sqrt{2}x_1x_2$, and $z_3 = x_2^2$. On the other hand, in the quantum-enhanced SVM, the embedding of data to $n$-qubit feature space can be performed by applying the unitary $\mathcal{U}_{\Phi(\ru{x})} =  \ru{U}_{\Phi(\ru{x})}\ru{H}^{\otimes n} \ru{U}_{\Phi(\ru{x})}\ru{H}^{\otimes n}$, where $\ru{H}$ is the Hadamard gate, and $\ru{U}_{\Phi(\ru{x})}$ denotes a diagonal gate in the Pauli-Z basis as below
\begin{eqnarray}
\ru{U}_{\Phi(\ru{x})} &=& \exp\left( i \sum_{S \subseteq \left[n\right]} \phi_S(\ru{x}) \prod_{k \in S} Z_k \right),\label{eq:vqc-real-mapping}
\end{eqnarray}
where the coefficients $\phi_S(\ru{x}) \in \mathbb{R}$ are fixed to encode the data $\ru{x}$. For example, for $n = d = 2$ qubits, $\phi_i(\ru{x}) = x_i$ and $\phi_{1,2}(\ru{x}) = (\pi - x_1)(\pi - x_2)$ were used in \cite{Havl_ek_2019}. In general the $\ru{U}_{\Phi(\ru{x})}$ can be any diagonal unitary that can be implemented efficiently with short-depth quantum circuits. In total, one needs at least $n \ge d$ qubits to construct such quantum-enhanced feature map. There are other proposed methods promising Quantum Advantage for non-linear embedding of data into quantum feature space, such as, \textit{squeezing} in continuous quantum systems \cite{Schuld_2019} that guarantees linear separability, or \textit{amplitude encoding} \cite{schuld2018circuitcentric} that can exploit tensorial feature map, or \textit{density-operator encoding} \cite{Mitarai_2018}. A recent paper studies the embedding in the context of \textit{metric learning}~\cite{lloyd2020quantum}. 

\begin{figure}[tb]
    \centering
    \includegraphics[width=0.9\columnwidth]{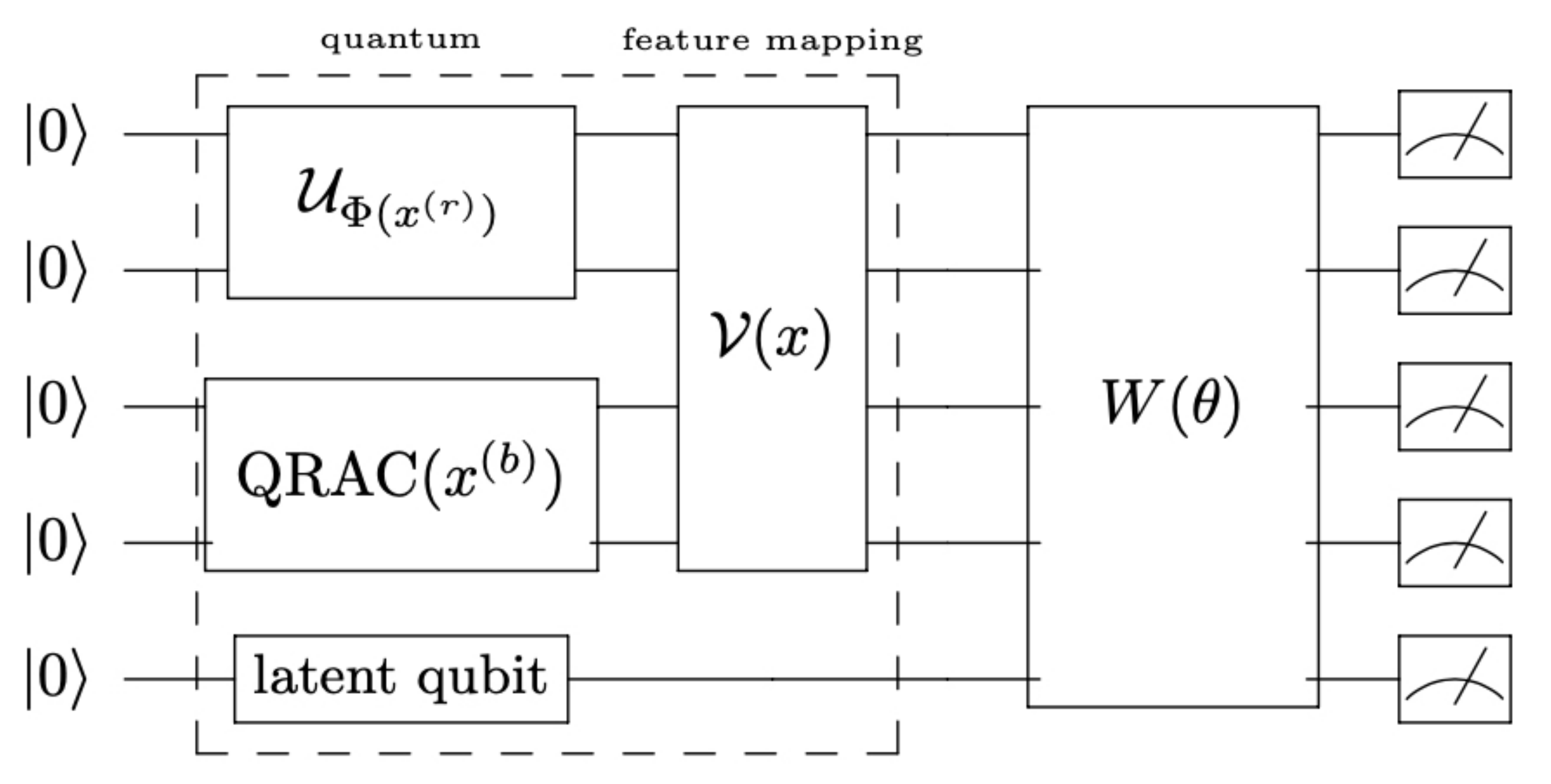}
    \caption{Quantum Circuits for Variational Quantum Classifier with Quantum Random Access Coding 
    for encoding discrete features. Latent qubits may be included to add the dimension of the embedding in the Hilbert space.}
    \label{fig:vqc-with-qrac}
\end{figure}

Classification models in real-world datasets often also depend on binary features, 
such as, gender and yes-no answers to questions, in addition to (discrete) categorical and qualitative features, 
such as, zip code, age, and color. Such discrete features have to be encoded into continuous features before 
they can be used effectively in machine learning models that rely on continuity of their inputs, such as VQCs. 
There have been many proposed encodings, with \textit{one-hot} encoding as one of the most populars, 
for such purposes~\cite{Hancock2020}. It is known that the encodings can heavily impact the performance 
of the learning models. Efficient mapping of such discrete features into quantum-enhanced feature space is very important in finance models with structured data. A recent study~\cite{YSRY2020} reports the possibility of using \textit{Quantum Random Access Coding}(QRAC)~\cite{ANTV99} to map discrete features into the quantum-enhanced feature space resulting in faster training and better classification accuracy due to using less number of qubits and hence less hyperparameters in the VQC models. The idea is to split the encoding of $\ru{x}$ into that for the discrete and continuous parts, each represented as $\ru{x}^{(b)}$, and $\ru{x}^{(r)}$. 
The discrete parts $\ru{x}^{(b)}$ are obtained from the encoding of categorical features 
into binary strings using determined techniques such as one-hot encoding, or into integer numbers for ordinal features. Fig.~\ref{fig:vqc-with-qrac} depicts a VQC with QRAC for encoding discrete features.

In particular, let us consider the case of classifying credit card transactions into fraudulent or not from a synthesized dataset from \cite{altman2019synthesizing}, that was generated with state machines in simulated world to be representative for the US customers. For our purpose, the synthesized credit card transaction data was prepared to contain 100 records of purchase transactions. The $i$-th transaction $\ru{x}_i$ contains the transaction time, the transaction amount, the transaction method, the transaction location (in ZIP code), and the Merchant Category Code (MCC). The first two are in real numbers, and the rest are categorical; there are 3 types of transaction methods, 10 different locations, and 10 different MCCs. Each $i$-th transaction is labeled as either fraudulent ($y_i = -1$), or normal ($y_i = 1$). 
A similar study on the same data set has also been carried out using \emph{Variational Quantum Boltzmann Machines}, an alternative approach to VQC or QKE, and we refer to \cite{Zoufal2020varqbm} for more details.

\begin{table*}[tb]
    \renewcommand{\arraystretch}{1.3}
    \caption{The average and standard deviation of accuracy of classifiers on 5-fold Cross Validations of the synthetic credit-card transaction dataset. LR, SVC1, and SVC2, are respectively, the Logistic Regression, the SVC with Linear, and RBF Kernel. VQC and VQCwQRAC are quantum-enhanced SVMs with the latter using QRAC for encoding the transaction method.}
    \label{tab:vqc-accuracy}
    \centering
    \sf
    \begin{tabular}{l r r r r r}
    \toprule
     &  LR & SVC1 & SVC2 & VQC &  VQCwQRAC\\
    \toprule
    train &$0.79 \pm 0.02$ & $0.80 \pm 0.03$& $0.87 \pm 0.02$& $0.74\pm0.05$ & $0.85\pm0.03$\\
    test  &$0.77 \pm 0.06$ & $0.81 \pm 0.07$& $0.82 \pm 0.07$& $0.64\pm0.09$ & $0.83\pm0.06$\\
    \bottomrule
    \end{tabular}
\end{table*}

\begin{figure}[tb]
    \centerline{\includegraphics[scale=0.35]{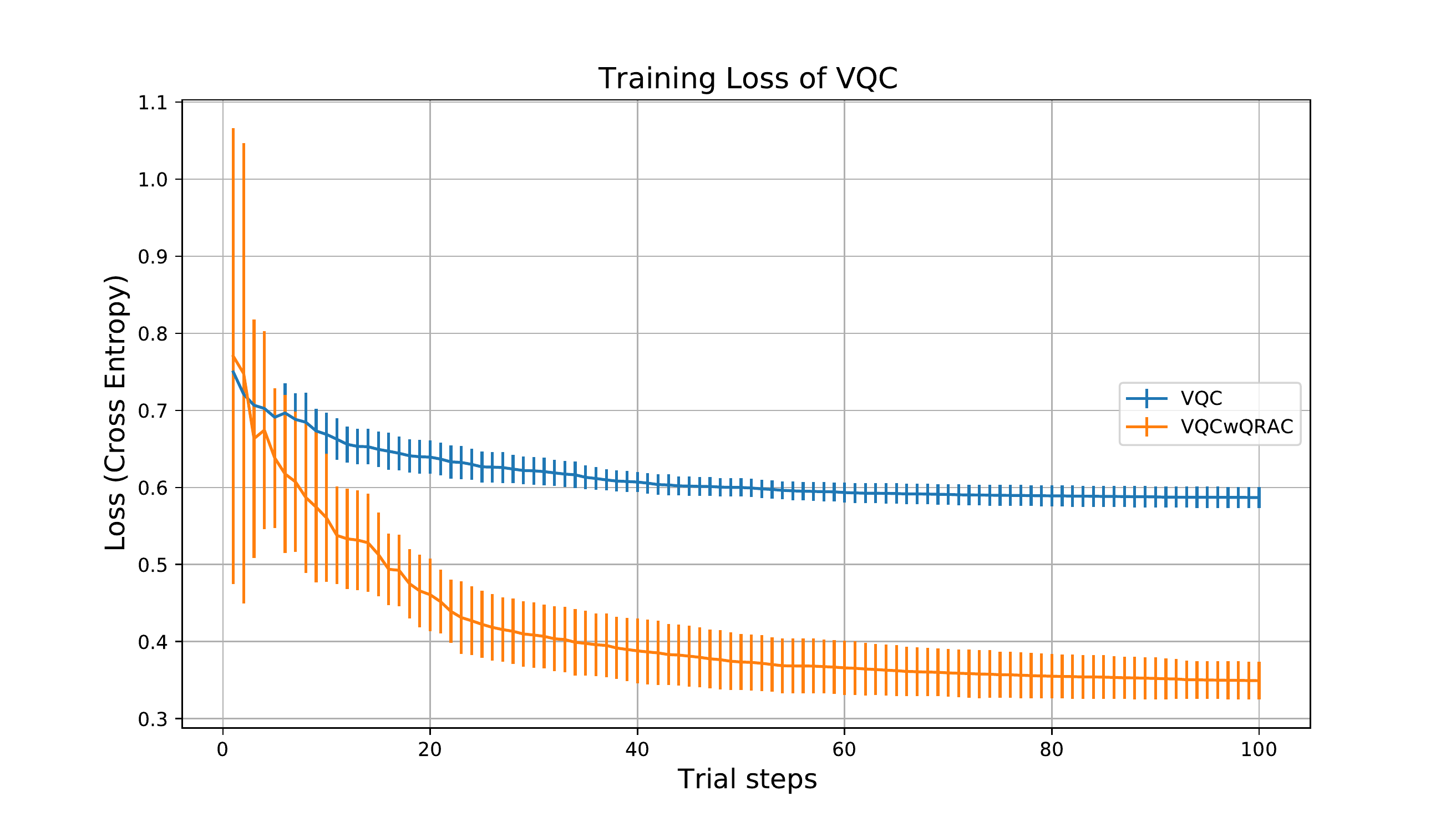}}
    \caption{Comparison of the training loss of VQC against that with QRAC (VQCwQRAC) for encoding discrete features on a synthetic credit-card transaction dataset.}
    \label{fig:vqc-loss}
\end{figure}

We applied the VQC as in Fig.~\ref{fig:vqc} by regarding all features as real values to use the feature mapping in Eq.~\ref{eq:vqc-real-mapping} with second-order expansion. On the other hand, we applied the VQC as in Fig.~\ref{fig:vqc-with-qrac} by the QRAC of the one-hot encoding of the transaction method, and the rest similar to the VQC. The latter is denoted as VQCwQRAC. Both models used 5 qubits and were trained with variational circuits $\ru{W}(\ru{\theta})$ defining the separating hyperplanes that consists of the RXRY variational gates and 1 layer of full-connected entangler as implemented in Qiskit\cite{Qiskit}. Both classification models were run on qiskit simulators and tested with 5-fold Cross Validation of the dataset. The average training losses, where Eq.~\ref{eq:emr} is approximated with the \textit{cross entropy}, are shown in Fig.~\ref{fig:vqc-loss}. We can see that using QRAC for encoding binary features can result in better training losses. The accuracy of VQCwQRAC is better than the VQC using real-valued quantum feature mapping as shown in Table~\ref{tab:vqc-accuracy}, and is comparable to Support Vector Classifier with RBF kernel (SVC2 in the Table).

Finally, we note that the possible Quantum Advantage for machine learning task is somewhat speculative; there is no known theoretical proof that the quantum feature map, which is hard to compute classically, can result in better accuracy than any classical classifiers. Also, the underlying variational methods, as in Algorithm~\ref{algo:VQE}~and~\ref{algo:QAOA}, are heuristics that may only find local optima instead of the global one, and thus can lower the accuracy of the resulting quantum-enhanced SVM.  

\section{Technical challenges in quantum computing}

In the following section we outline some of the technical challenges to address when solving computationally challenging problems on a quantum computer.

\subsection{Loading data}
\label{sec:dataloading}

To understand constraints on quantum computing both near-term and long-term, it may be useful to contrast quantum computers against classical computers.
Classical computing utilises the well-known von Neumann model: there is a central processing unit (CPU), which performs non-reversible computation, including branching, and this is connected by a system bus to volatile memory (RAM) and non-volatile memory (such as a hard drive). Loading data from the non-volatile memory to RAM and accessing the data in RAM from the CPU is taken for granted. 
In contrast, there are no quantum (memory) hard drives at the current level of hardware technology, most blueprints do not involve any RAM, and all of the computations are reversible without branching (excepting post-selection).

The key difference lies in the time complexity of ``loading data''. 
A quantum state can be seen as a volatile memory of substantial capacity, but with non-trivial issues in addressing it.
With $k$ qubits, we work with $2^k \times 2^k$
density matrices \footnote{A density matrix is a representation of a quantum state. A density matrix can represent both pure states (i.e. states represented by a state vector $\ket{\psi}$) and mixed quantum states (i.e. a statistical ensemble of pure states).},
but working with these matrices is limited to a certain set of one- and two-qubit gates (unitary matrices applied to
the quantum state). 
The quantum circuit complexity of state preparation, i.e.,
minimum number of gates required in order to ``load'' a given \rev{arbitrary} quantum state $U$ 
using {\em any} sequence of one- and two-qubit gates, 
is greater or equal than $4^k$ for almost all $U$.
Notice that this is not a worst-case result: this holds generically for all states 
and it applies to the best possible sequence of one- and two-qubit gates.
Consider a dimension-counting argument. There are also explicit constructions showing that this is tight.

\rev{
While this means theoretically that a $4^k$-dimensional state can be prepared into a quantum machine having $k$ qubits
via a quantum loader having $O(4^k)$ circuit complexity --
which is linear in the data dimension -- it is also 
exponential in the number of qubits $k$.
Whereas in classical computing, we usually assume that we can load the data in
time linear in the number of bits
and then worry about the run-time (circuit complexity of processing) on the loaded data,
in quantum computing, preparation of a quantum state may require  
a quantum circuit complexity exponential in the number of qubits.
The complexity of generic state preparation can impede a potential Quantum Advantage, as the loading time may eclipse coherence times
for the physical quantum state and also for some algorithms,
loading the data can become computationally as expensive as using a classical algorithm to solve the problem \cite{CHIPRSW2018}.
}

There are multiple ways of circumventing this issue.
One is to allow $j$-qubit gates, where $j$ may grow with $k$,
which poses a major challenge in quantum optimal control.
One is to ``split'' the state preparation into an independent system, 
such as circuitry of some potential qRAM \cite{Jiang2019},
and utilise quantum optimal control there, perhaps across all of the $k$ qubits.
Given the (quantum circuit complexity) equivalence \cite{nielsen2006optimal} of state preparation and an arbitrary circuit, it seems unlikely that it would be possible to implement one way of circumventing the quantum circuit complexity without being able to implement the other, and without being able to utilise the same quantum optimal control in the execution of the quantum circuit.
\rev{Indeed}, it is believed the physical realization of qRAM model may be even much more difficult than the fault-tolerant quantum computers \cite{Biamonte2017,CHIPRSW2018}. 

In some cases, the problem can also be circumvented because the data to be loaded has structure or properties that can be exploited for efficient loading, e.g., if the data can be described by a log-concave probability distribution \cite{grover2002logconcave}.
Alternatively, and depending on the application, we may drop the goal of loading the data exactly and try to prepare a quantum state that at least is close to our original data. 
This enables approximate data loading schemes which have some potential to work around this problem \cite{zoufal2019qgan}. It is also possible to exploit periodic properties in certain data sets, for example. time series data which exhibits periodic properties. By extracting periods in the data via classical techniques such as FFT, we may then load only the dominant periods via a small number of steps onto the quantum machine, then recover an approximation of the original data in the quantum machine via an inverse quantum Fourier transform (QFFT$^{-1}$) algorithm.

\subsection{Quantum error correction}

As has been suggested in Section \ref{intro}, a key watershed between noisy quantum computers and universal fault-tolerant quantum computers is the availability of quantum error correction (QEC).

The key technical challenge within QEC is the trade-off between the overhead of the QEC and the so-called error threshold.
The overhead is, essentially, the \rev{number} of physical qubits required to protect a certain number of logical qubits against errors. 
The error threshold comes from the famous (quantum) threshold theorem \cite{gottesman1997stabilizer,aharonov2008fault}, which shows that if the 
errors on individual qubits are not correlated and the 
error of the physical qubits falls below a certain threshold, QEC schemes can correct the remainder of the error, at a cost of the overhead.
Actually, the dissertation of Gottesman \cite{gottesman1997stabilizer} shows that there is a simple construction, starting with classical error correcting codes, which makes it possible to estimate the threshold.
For one of the best-known classes of QEC, it is sometimes assumed \cite{campbell2017roads} that 
a 0.1\% probability of a depolarizing error would require more than 1,000 physical qubits per protected qubit -- although the details of the calculation are also often disputed. 
There is a substantial interest in further classes of QEC (e.g., hyperbolic surface codes),
which could perform substantially better. 

For the same QEC mentioned above \cite{campbell2017roads}, one should notice that there need not be a substantial increase in the depth of the circuit: for gates within the Clifford algebra, which includes the Hadamard gate (H), controlled not (CNOT), and $S = \textrm{diag}(1, e^{i\pi/2})$, we can apply the gate to all the physical qubits in order to apply the same gate to the protected qubit. 
The increase in depth of the circuit is hence only due to gates outside of the Clifford algebra.

\rev{In Sec.~\ref{sec:simulation} we showed that AE can replace Monte Carlo based simulations.
The resulting quantum circuits are too deep for quantum computers without error correction due to the controlled $\mathcal{Q}^{2^j}$ operators, see Fig.~\ref{fig:amplitude_estimation_circuit}.
We therefore anticipate that AE-based applications will require fault-tolerant quantum computers.
By contrast, the optimization and machine learning applications discussed in Sec.~\ref{sec:optimization} and \ref{sec:ml} that are based on variational quantum circuits as in the VQE and the QAOA could be executed on near-term noisy quantum computers.
However, these heuristic algorithms do not provide a theoretical guarantee as does AE.
Further research is thus needed to fully understand under what conditions they will outperform their classical counterparts.}

\subsection{Precision and sample complexity}
\label{sec:samplecomplexity}

Generally, the higher probability of outputting the correct answer is required,
the more ``shots'', or repetitions of the execution of quantum circuit followed by measurement, are needed.
In some cases (e.g., HHL), because the solution is encoded in the probability amplitudes of the quantum states, one may need to perform quantum state tomography to obtain the complete solution. The quantum state tomography requires exponential number of shots in the number of qubits involved, and hence can diminish the exponential advantages of the subroutines.
In many algorithms, the error also depends on the number of qubits used in the output register.

For example for the phase estimation mentioned in Section \ref{sec:ae},
the probability of not determining phase angle to an accuracy of $s$ bits, 
i.e., up to error $2^{-s}$, using $s + p$ qubits for the output is:
\begin{align}
\epsilon(s, p) = 1 - \frac{1}{2^{2(p+s)-2}} \sum_{l = 1}^{2^{p-1}}\frac{1}{1 - \cos \frac{\pi(2l - 1)}{2^{p+s}} }
\end{align}

While the expression may be difficult to read, it is essentially positive, in that it suggests that the error rate decays exponentially with the 
extra $p$ qubits.
Especially when many instances of phase estimation are used sequentially, the error propagation may still be a cause for concern, though,
and it may get progressively more difficult to analyze the error.
Still, estimates of forward error of more complex algorithms \cite{Cleve1998}
are available.

\begin{table*}
\sf
\caption{Algorithms can improve computational efficiency, accuracy, and addressability for defined use case.}
 \begin{adjustbox}{max width=\textwidth}
\begin{tabularx}{800pt}{lXXXXccc}
\toprule
& Quantum Algorithm & 
Description &  
Impact & 
Needs & 
\rotatebox[origin=c]{90}{Simulation} & 
\rotatebox[origin=c]{90}{Optimization} &  
\rotatebox[origin=c]{90}{ML} \\
\toprule
VQE & Variational Quantum Eigensolver & 
Use energy states to calculate the function of the variables to optimize & 
Procedure to assign compute-intensive functions to quantum and those of controls to classical &
Qubit number increases significantly with problem size & 
 & 
x & 
\\[2em] \midrule
QAOA & Quantum Approximate Optimization  & 
Optimize combinatorial style problems to find solutions with complex constraints  & 
Simplify analysis clauses for constraints and provide robust optimization in complex scenarios  & 
Uncertain ability to expand to more optimization classes & 
 & 
x & 
\\[2em] \midrule
\rev{AE} & Quantum Amplitude Estimator & 
Create simulation scenarios by estimating an unknown property, Monte Carlo style & 
Handle random distributions directly, instead of only sampling, to solve dynamic problems quadratically speeding up simulations & 
High Quantum Volume required for good efficiency & 
x  & 
x  & 
x \\[2em] \midrule
QSVM & Quantum Support Vector Machines  & 
Supervised machine learning for high dimensional problem sets  & 
Map data to  quantum-enhanced feature space to enable separation and better separate data points to achieve more accuracy & 
Runtime can be slowed by kernel computation and data structure  & 
 & 
 & 
x\\[2em] \midrule
HHL & Harrow, Hassidim, and Lloyd  & 
Estimate the resulting measurement of large linear systems  & 
Solve high dimensional problems speeding up exponentially calculations  & 
Hard to satisfy prerequisites and high measurement costs to recover solutions & 
 & 
x & 
 x\\[2em] \midrule
QSDP & Quantum Semidefinite Programming  & 
Optimize a linear objective over a set of positive semidefinite matrices & 

Estimate quantum system states with less measurements to exponentially speedup in terms of dimension and constraints  & 
High Quantum Volume required for good efficiency & 
 & x
 & \\
\bottomrule
\end{tabularx}
\end{adjustbox}

\label{tab1}
\end{table*}

\begin{table}
\caption{Financial services focus areas and algorithms.}
\sf
 \begin{adjustbox}{max width=\columnwidth}
\begin{tabularx}{400pt}{lllX}
\toprule

Financial Services & Example Problems & Solution Approach & Quantum Algorithm\\
\toprule
Asset Management & Option Pricing & Simulation & \rev{AE} \\
& Portfolio risk & & \\
\midrule
Investment & Portfolio Optimization & Optimization & Combinatorial: VQE, QAOA\\
Banking & Portfolio Diversification & & Continuous: QSDP \\
& Issuance: Auctions & & \rev{AE}\\
\midrule
Retail \& Corporate  & Financial Forecasting & Machine Learning & QSVM \\
Banking & Credit Scoring (e.g. SME Banking) & & HHL \\
& Financial Crimes: Fraud + AML &  & \rev{AE} \\
\bottomrule
\end{tabularx}
\end{adjustbox}
\label{tab2}
\end{table}

\section{Conclusion}

There are a number of computationally challenging problems in financial services that are demanding in terms of required precision or run-time. For these we have outlined three problem classes:

\begin{itemize}
\item
One class are \textit{optimization} problems that scale exponentially limiting their resolution in a given time frame. The holistic problem-solving approach to optimization problems of quantum computers, raises the potential to find better solutions in a smaller number of steps. 
\item
A second class are \textit{machine learning} problems, where one faces complex data structures that hinder classification or prediction accuracy. The multi-dimensional data modeling capacity of quantum computers may allow to find better patterns, with increasing accuracy.
\item
A third class are \textit{simulation} problems, where there are time-limits to perform sufficient scenario tests to find the best potential solution. Efficient sampling methods leveraging quantum computers may require less samples to reach a more accurate solution faster.

\end{itemize}

For each of them we have introduced quantum algorithms, which can be applied to specific problems in there. Table \ref{tab1} summarizes the quantum algorithms introduced, their applicability for the three problem classes, their advantages and challenges. In addition, for the three initial focus areas in financial services, Asset Management, Investment Banking, Retail \& Corporate Banking, example problems and applicable quantum algorithms are summarized in Table \ref{tab2}.

Quantum computers and the algorithms that leverage them may help to solve hurdles and challenges arising in the financial industry given  increasing demand for more sophisticated risk analysis, dynamic client management, constant updates to market volatility, and faster transaction speeds. \\

Finally, we have also demonstrated the performance of quantum algorithms on IBM Quantum backends for three specific applications. In general, simulation, optimization and machine learning are among the areas where we may demonstrate an advantage of quantum computing over classical computing for certain applications first.

\section*{Acknowledgements}
RR would like to acknowledge Erik Altman of IBM Research for providing the credit-card transaction dataset, and Hiroshi Yano of Keio University for his help in analyzing the dataset. 

IBM, the IBM logo, and ibm.com are trademarks of International Business Machines Corp., registered in many jurisdictions worldwide. Other product and service names might be trademarks of IBM or other companies. The current list of IBM trademarks is available at \url{https://www.ibm.com/legal/copytrade}.
\newpage 

\smallskip
\bibliographystyle{elsarticle-num} 
\bibliography{finance} 

\EOD

\end{document}